\documentclass[12pt]{article}
\usepackage{jheppub}

\def\CL{{\cal L}}

\title{Open Verlinde line operators}

\author[1]{Davide Gaiotto}

\affiliation[1]{Perimeter Institute for Theoretical Physics\\%
31 Caroline Street North, ON N2L 2Y5, Canada}

\abstract{We reformulate the action of Verlinde line operators on conformal blocks in a 3d TFT language and extend it to 
line operators labelled by open paths joining punctures on the Riemann surface. We discuss the possible applications of open Verlinde line operators to 
quantum Teichm\"uller theory, supersymmetric gauge theory and quantum groups}

\begin{document}
\maketitle
\section{Introduction}
Verlinde line operators are a class of linear operators which act on the space of conformal blocks for some current algebra \cite{Verlinde:1988sn}, 
which in this note will always coincide with the Virasoro algebra. They are roughly labeled by a closed path $\ell$ on the Riemann surface, up to some subtleties which we will review momentarily. Traditionally, are defined by composing a sequence of elementary operations on conformal blocks, each corresponding to a map between spaces of conformal blocks which may differ in the number or type of insertions.
Roughly speaking, one inserts an identity operator into the original conformal block, splits into two conjugate chiral operators
$\phi_a$ and $\phi_{\bar a}$, transports $\phi_a$ along $\ell$ and then fuses the operators $\phi_a$ and $\phi_{\bar a}$ back to the identity channel.

Although originally defined in the context of 2d RCFTs, the definition works well for general Virasoro conformal blocks and 
give operators with interesting interpretations in the context of supersymmetric gauge theory \cite{Drukker:2009id,Alday:2009fs,Gomis:2011pf}.

This definition has some rough edges. The answer may depend on how the path $\ell$ winds around the auxiliary insertion $\phi_{\bar a}$
and it is not obvious that it will be independent on the choice of starting point along $\ell$. 
Typically, the normalization of the operator has to be set by hand. Experimentally, the Verlinde line operators satisfy nice skein relations, 
which again are not quite obvious from the definition and seem to imply a 3d TFT interpretation: 
the operators should really be labelled by paths drawn close to the Riemann surface into a three-dimensional geometry bounded by the surface itself.

An important observation is that Verlinde line operators are closely related to topological defects in CFTs. More precisely, if a CFT correlation function is 
represented as a pairing between anti-holomorphic and holomorphic conformal blocks, a Verlinde line operators inserted in the pairing should modify 
the correlation function precisely in the same way as a topological line defect with the same Cardy label as the 
original chiral operator $\phi_a$ \cite{Drukker:2010jp}. This is at least true both in RCFTs and in almost-rational CFTs such as Liouville theory. 

Topological line defects in RCFTs do have a neat 3d TFT interpretation \cite{Fjelstad:2012mj} in terms of 
graphs embedded into a three-dimensional manifold, bound by two copies of the Riemann surface, 
where chiral and anti-chiral operators live. A secondary objective of this note will be to 
give a similar 3d TFT-inspired alternative definition of Verlinde line operators. 
The three-dimensional geometry will be a handle-body bounded by the Riemann surface, as appropriate 
for describing conformal blocks. 

In the context of RCFTs, one can consider both topological defects wrapping closed paths and topological defects wrapping open paths ending at 
punctures. The main purpose of this note will be to extend the definition of Verlinde line operators to encompass operators 
labeled by open curves, which end at the punctures of the conformal block. Our definitions can be found in section \ref{sec:verlinde}
and concrete calculations in sections \ref{sec:for} and \ref{sec:ex}.

We will also describe briefly three applications of open Verlinde line operators, leaving a complete analysis to later publications. 
The first one is to establish a direct connection from the theory of BPZ conformal blocks to quantum Teichm\"uller theory: 
we will recover the quantum version of Fock coordinates as cross-ratios of open Verlinde line operators.
This allows one to reverse the logic of standard constructions of quantum Teichm\"uller theory such as \cite{2005math}.
The second is to define the quantum deformation of the traffic rules used in computing framed BPS degeneracies for theories of the class $A_1$ \cite{Gaiotto:2010be}.
The third is to clarify the relation between BPZ conformal blocks and quantum groups, and the identity between the fusion integral kernels and the quantum 6j symbols \cite{2013arX}. 
To do so, we will use a construction originally due to Witten in the context of Chern-Simons theory \cite{Witten:1989rw}. All applications are discussed in section \ref{sec:app}.

\section{A 3d perspective}

A standard conformal block can be pictured as a trivalent graph $\Gamma_0$, representing a pair of pants decomposition of the Riemann surface.
The graph should be thought of as embedded in the middle of the 3d handle-body associated to that decomposition, 
which fills in the Riemann surface by making the tubes of the pair of pants into solid cylinders. 
In genus zero, which will be our main source of examples, $\Gamma_0$ is a trivalent tree graph drawn inside a 3d ball, 
with external legs ending at the boundary of the ball. 

\begin{figure}
\center \includegraphics[height=4cm]{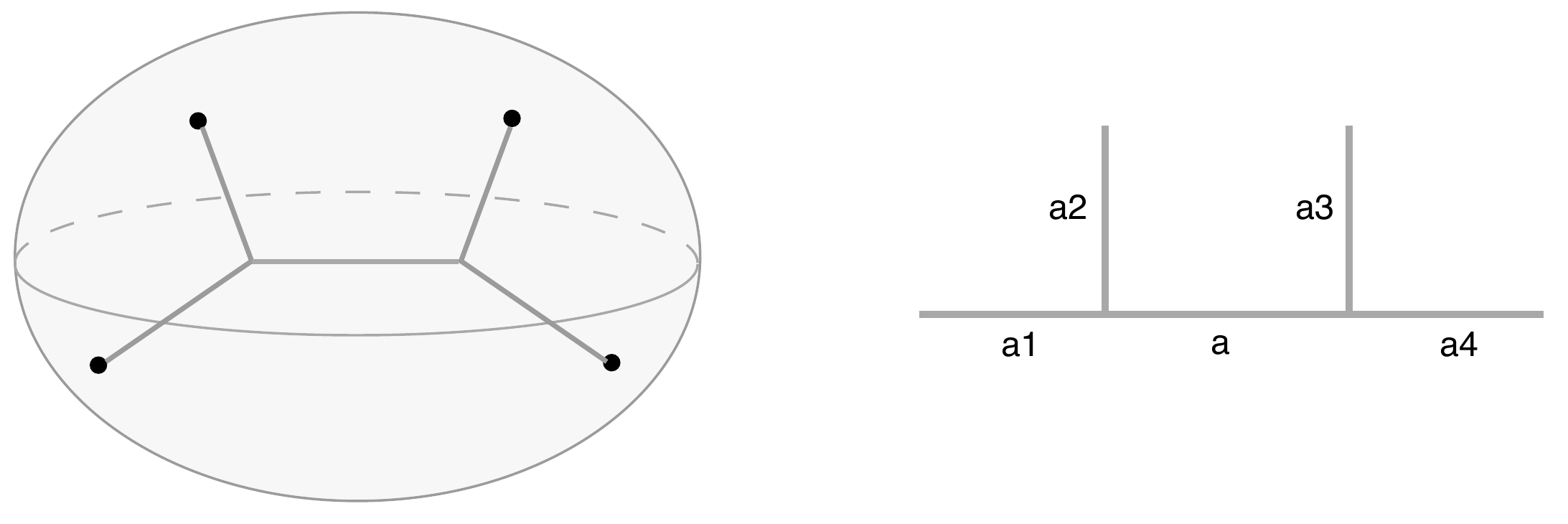}
\caption{\label{fig:4pt} Left: the pictorial representation of a four-point conformal block, as a trivalent graph inside a ball. We omit for clarity the Liouville momentum labels on the edges. Right: The same conformal block, with the 3d ambient space omitted, and Liouville momenta labels on the edges.}
\end{figure}

Each edge of the graph is labeled by a representation of the Virasoro algebra, usually denoted by a choice of Liouville momentum. 
See Figure \ref{fig:4pt} for an example. 
In this note we will encounter both generic representations and degenerate representations of the Virasoro algebra. We will 
always impose the decoupling of null vectors 
for the degenerate representations, which implies specific selection rules on the Liouville momenta which flow through the graph. 

We can denote a conformal block associated to the trivalent graph $\Gamma_0$ in a ket notation
\begin{equation}
|a_i\rangle_{\Gamma_0}
\end{equation}
with $a_i$ being the Liouville momenta on the edges. 
We will typically leave implicit the dependence on the complex structure parameters on the Riemann surface,
and omit the $\Gamma_0$ subscript when unambiguous.

\begin{figure}
\center \includegraphics[height=4cm]{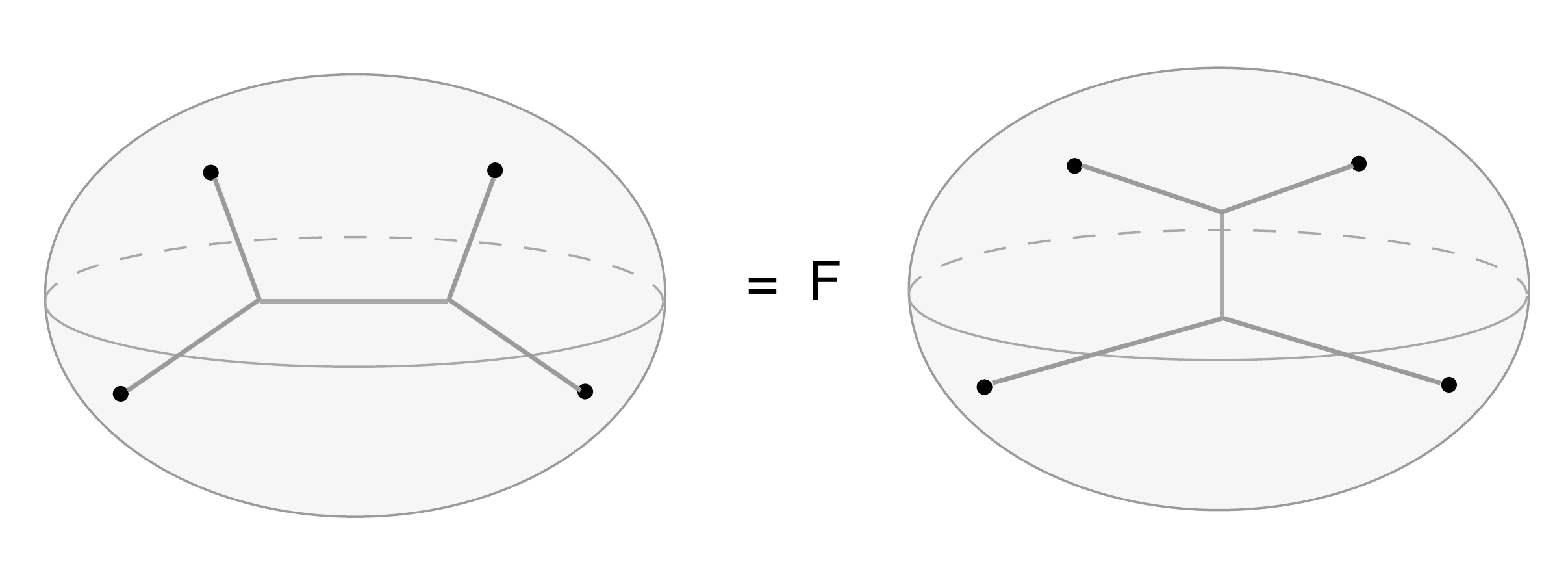}
\caption{\label{fig:4ptF} The pictorial representation of a fusion operation on the four-point conformal block. The fusion ``matrix'' $F$ acts on the Liouville momentum label on the internal edge. It will be an integral kernel if the external edges carry a generic Liouville momentum, and a finite matrix if at least one external legs is degenerate}
\end{figure}

Conformal blocks associated to different pair of pants decompositions are related by linear transformations. 
Different pair of pants decompositions can be related by a sequence of elementary 
fusion and braiding moves (and the $S$ transformation in positive genus), which are associated to well-known linear transformations.  
See Figure \ref{fig:4ptF} for the pictorial representation of a fusion operation. A fusion operation which involves non-degenerate representations only 
is associated to an integral transformation with complicated integration kernel, as the s-channel conformal block is expressed as an integral over all possible t-channel Liouville momenta. If at least one of the representations involved is degenerate, then one deals with a much simpler, finite-dimensional linear transformation. 

The standard definition of Verlinde line operators, and their open generalizations, involves a formal sequence of such fusion and braiding transformations.
Our first step will be to introduce some 3d TFT-inspired definitions which help give a precise meaning to the intermediate steps of the calculation of 
Verlinde line operators. 

For simplicity, we will restrict ourselves to situations where we encounter only the finite-dimensional fusion transformations associated to degenerate fields. 
Our abstract definitions could be generalized to the infinite-dimensional case in a straightforward way, but computations would be much harder. 

\subsection{Generalized conformal blocks} \label{sec:verlinde}
In order to do 3d TFT-like manipulations on conformal blocks, we will define a map from a much larger class of trivalent graphs in a fixed handle-body to the space of conformal blocks associated to a given pair of pants decomposition. The graphs may include many more loops and connected components than the basic trivalent graph $\Gamma_0$, but 
live in the handlebody associated to $\Gamma_0$. 

The graphs must be refinements of the standard trivalent graph $\Gamma_0$ associated to the handle-body.
More precisely, any such graph $\Gamma$ must satisfy the following constraints:
\begin{itemize}
\item Each edge $e_i^0$ of $\Gamma_0$ will map to a consecutive sequence of edges $e_{i,u}$ in $\Gamma$. 
\item The Liouville momenta on each collection of edges $e_{i,u}$ of $\Gamma$ must differ by integral multiples of $\i b/2$ and $\i b^{-1}/2$ 
from the common value $a_i$ associated to $e_i^0$ in $\Gamma_0$: 
\begin{equation} a_{i,u} = a_i + \i s_{i,u}\frac{b}{2} + \i \tilde s_{i,u} \frac{b^{-1}}{2}\end{equation}
\item All other edges of $\Gamma$ must carry degenerate representations. 
The graph $\Gamma$ may include closed loops disconnected from the rest of the graph. 
\item At all vertices involving degenerate edges, the degenerate fusion rules must hold. \end{itemize}

We will typically include the choice of degenerate representations in the new edges and 
of the $s$ and $\tilde s$ shifts implicitly in the label ``$\Gamma$''. If we want to make the shifts explicit, we may refer to 
a graph $\Gamma[s,\tilde s]$. In concrete examples, our lines will be of type $(r,1)$, and thus we will only need shifts $s$ by multiples of $\i b/2$,
no $\tilde s$ shifts. Whenever we use the term ``graph'' from now on, we mean to denote such a decorated refinement of $\Gamma_0$. 

\begin{figure}
\center \includegraphics[height=4cm]{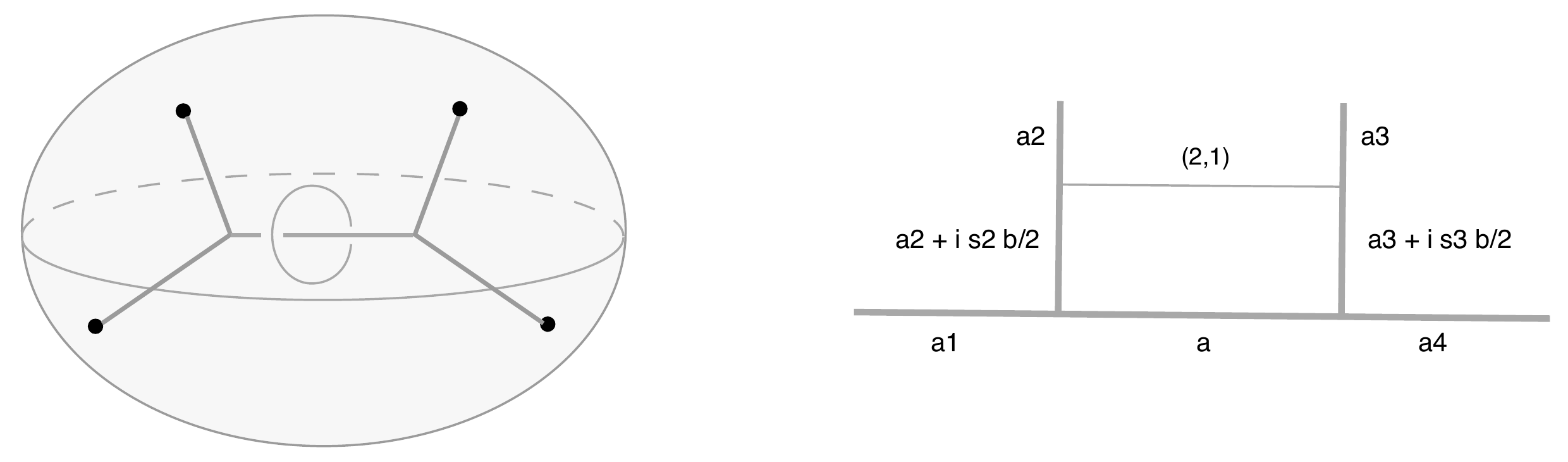}
\caption{\label{fig:4ptdec} Left: a graph which refines the basic four-point graph $\Gamma_0$ in figure \ref{fig:4pt} by an extra degenerate loop linking the middle edge of $\Gamma_0$. We denote the degenerate loop by a thinner line. Right: a graph which refines the basic four-point graph $\Gamma_0$ by an extra rung. We omit the ambient three-ball and indicate the Liouville momenta. The $(2,1)$ label indicates the $\i b + \i b^{-1}/2$ degenerate momentum. The shifts $s_2$, $s_3$ take values $\pm1$}
\end{figure}

We can describe two simple examples of graphs $\Gamma$, depicted in Figure \ref{fig:4ptdec}: 
\begin{itemize}
\item The union of $\Gamma_0$ and one or more closed loops linked to it
\item A graph $\Gamma$ obtained from $\Gamma_0$ by inserting an extra rung, a degenerate edge attached to two edges of $\Gamma_0$.
Each end of the extra edge splits the corresponding edge of $\Gamma_0$ into two parts, with momenta which differ by an amount compatible with degenerate fusion. 
\end{itemize}

 The conformal block associated to a given graph $\Gamma$ will be a finite linear combination of conformal blocks associated to $\Gamma_0$, with Liouville momenta shifted by appropriate multiples of $b$ and $b^{-1}$:
\begin{equation}
|\Gamma;a_i\rangle_{\Gamma_0} = \sum_{n_i, \tilde n_i} c^{\Gamma_0;\Gamma}_{n_i, \tilde n_i} |a_i + n_i \frac{b}{2} + \tilde n_i \frac{b^{-1}}{2}\rangle_{\Gamma_0}
\end{equation}
The sum will never include shifts of the Liouville momenta on external edges. By definition, $|\Gamma_0;a_i\rangle_{\Gamma_0} = |a_i\rangle_{\Gamma_0} $.

The coefficients $c^{\Gamma_0;\Gamma}_{n_i, \tilde n_i}$ will be rational functions in $\exp \pi b a_i$ and $\exp \pi b^{-1} a_i$ 
We can re-write the above relation in operatorial form, defining Weyl-commuting operators
\begin{align}
&\hat \alpha_i |a_i\rangle_{\Gamma_0} = \exp \pi b a_i |a_i\rangle_{\Gamma_0} \qquad \qquad &\hat \alpha'_i |a_i\rangle_{\Gamma_0} = \exp \pi b^{-1} a_i |a_i\rangle_{\Gamma_0} \cr &\hat p_i |a_i\rangle_{\Gamma_0} = |a_i + \i b/2 \rangle_{\Gamma_0} \qquad \qquad &\hat p'_i |a_i\rangle_{\Gamma_0} =  |a_i + \i b^{-1}/2 \rangle_{\Gamma_0} 
\end{align}
and thus 
\begin{equation}
|\Gamma;a_i\rangle_{\Gamma_0} = \hat O_{\Gamma} |a_i \rangle_{\Gamma_0} \qquad \hat O_{\Gamma} = \sum_{n_i, \tilde n_i} c^{\Gamma_0;\Gamma}_{n_i, \tilde n_i}(\hat \alpha, \hat \alpha') \hat p_i^{n_i} (\hat p'_i)^{\tilde n_i} 
\end{equation}
We will drop the hats when possible to do so unambiguously to lighten the notation. Occasionally, we will also use the un-exponentiated operator 
\begin{equation}
\hat a_i |a_i\rangle_{\Gamma_0}  = a_i |a_i\rangle_{\Gamma_0} 
\end{equation}

The definition of these conformal blocks is recursive: if any pair of of graphs $\Gamma$ and $\Gamma'$ (which refine the same $\Gamma_0$) 
are related by a fusion or braiding move involving one of the extra degenerate edges, then the corresponding conformal blocks $|\Gamma;a_i\rangle_{\Gamma_0}$ 
and $|\Gamma';a_i'\rangle_{\Gamma_0}$ are related by the corresponding linear relationship. 

Furthermore, we are allowed to freely add or remove any number of rungs labeled by the identity representation, without changing the corresponding conformal blocks. If any subgraph is contained in a three-ball, whose boundary is crossed by a 
single edge which carries a representation different from the identity, the corresponding conformal block is zero. 
The final axiom is that a contractible loop labeled by a degenerate representation can be removed, multiplying the conformal block by the quantum dimension of the degenerate representation.  

In order to compute $|\Gamma;a_i\rangle_{\Gamma_0}$, one simply applies a sequence of such moves to $\Gamma$, 
until it is reduced to $\Gamma_0$. 
 
A good way to gain intuition on our construction is to depict conformal blocks as the result of a calculation in some 3d TFT, such as some kind of Chern-Simons theory 
\footnote{It is unclear if such a 3d TFT truly exists for general conformal blocks. The closest to it is likely some construction involving a twisted ${\cal N}=4$ SYM theory \cite{Witten:2011zz,Gaiotto:2011nm}.},
involving the handle-body geometry and a network $\Gamma_0$ of line defects. The space of conformal blocks is the Hilbert space associated to 
the boundary of the geometry, and the partition function with the network $\Gamma_0$ produces the specific basis element $|a\rangle$. 
Our $|\Gamma;a_i\rangle_{\Gamma_0}$ then should represent the answer to a 3d TFT calculation involving the same 3d geometry and 
the graph $\Gamma$ of line defects. We set up the problem in such a way that $\Gamma$ can be reduced to $\Gamma_0$ 
by using only 3d TFT manipulations involving degenerate lines.

 \subsection{Verlinde line operators} 
\begin{figure}
\center \includegraphics[height=3.5cm]{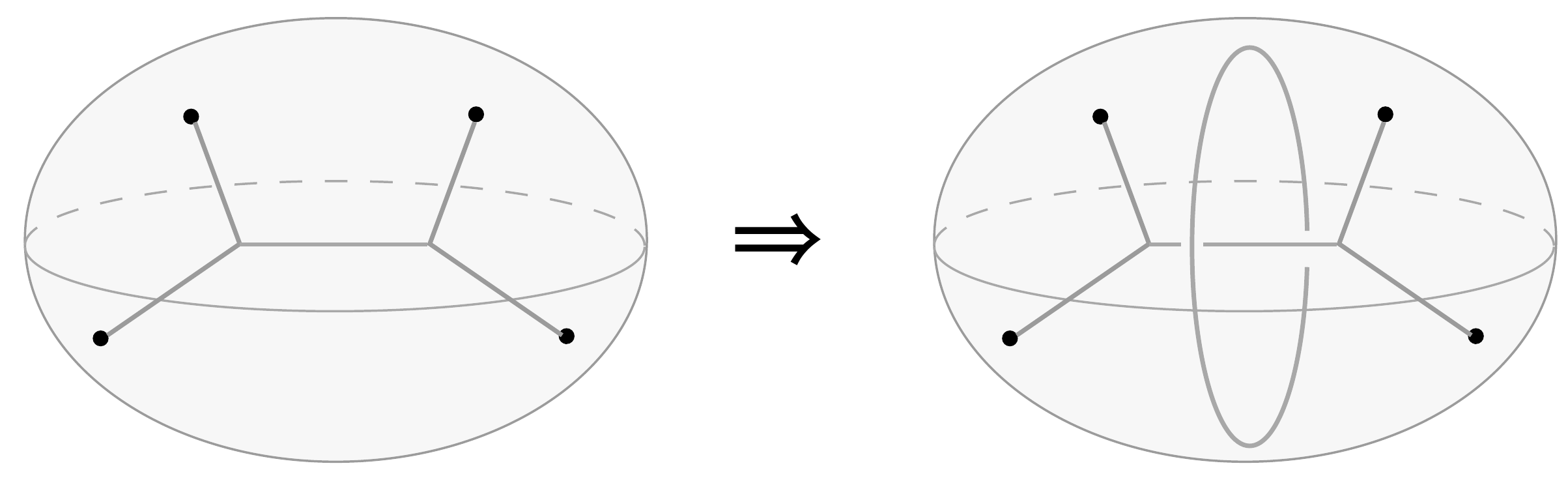}
\caption{\label{fig:4ptAloop} The pictorial representation of an ``A-cycle'' Verlinde line operator acting on 
a four-point conformal block. We omit for clarity the Liouville momentum labels on the edges}
\end{figure}

The action of a closed Verlinde line operators on a conformal block can be pictured as the insertion in the 3d manifold of an extra closed line, 
disconnected from the original graph, which runs just below the boundary along the path which labels the line operators. 
See Figure \ref{fig:4ptAloop} for an example. 
The new line carries a specific degenerate Virasoro representation which also characterizes the choice of line defect. 
We can denote such a line, together with the choice of representation, with the symbol $\ell$.

Given a graph $\Gamma$, which may or not coincide with the standard $\Gamma_0$, the addition of the extra line $\ell$ 
will produce a new graph $\ell \circ \Gamma$. The Verlinde line operator is a shift operator $\hat O_\ell$ defined as 
\begin{equation}
\hat O_\ell \equiv \hat O_{\ell \circ \Gamma_0}
\end{equation}
It is easy to see it satisfies the relation:
\begin{equation}
|\ell \circ \Gamma;a_i\rangle_{\Gamma_0} = \hat O_{\ell} |\Gamma;a_i\rangle_{\Gamma_0}
\end{equation}
Indeed, we can do the sequence of manipulations which reduce $\Gamma$ to $\Gamma_0$ 
first inside $\ell \circ \Gamma$, and then do the manipulations which reduce $\ell \circ \Gamma_0$ to $\Gamma_0$. 

In particular, 
\begin{equation}
|\ell_1\circ \ell_2 \circ \Gamma;a_i\rangle_{\Gamma_0} = \hat O_{\ell_1}\hat O_{\ell_2} |\Gamma;a_i\rangle_{\Gamma_0}
\end{equation}
etcetera. 
Operators labelled by lines which do not intersect when projected on the Riemann surface will commute, and the lines can be slipped through each other in the 3d geometry. 
 
\begin{figure}
\center \includegraphics[height=6cm]{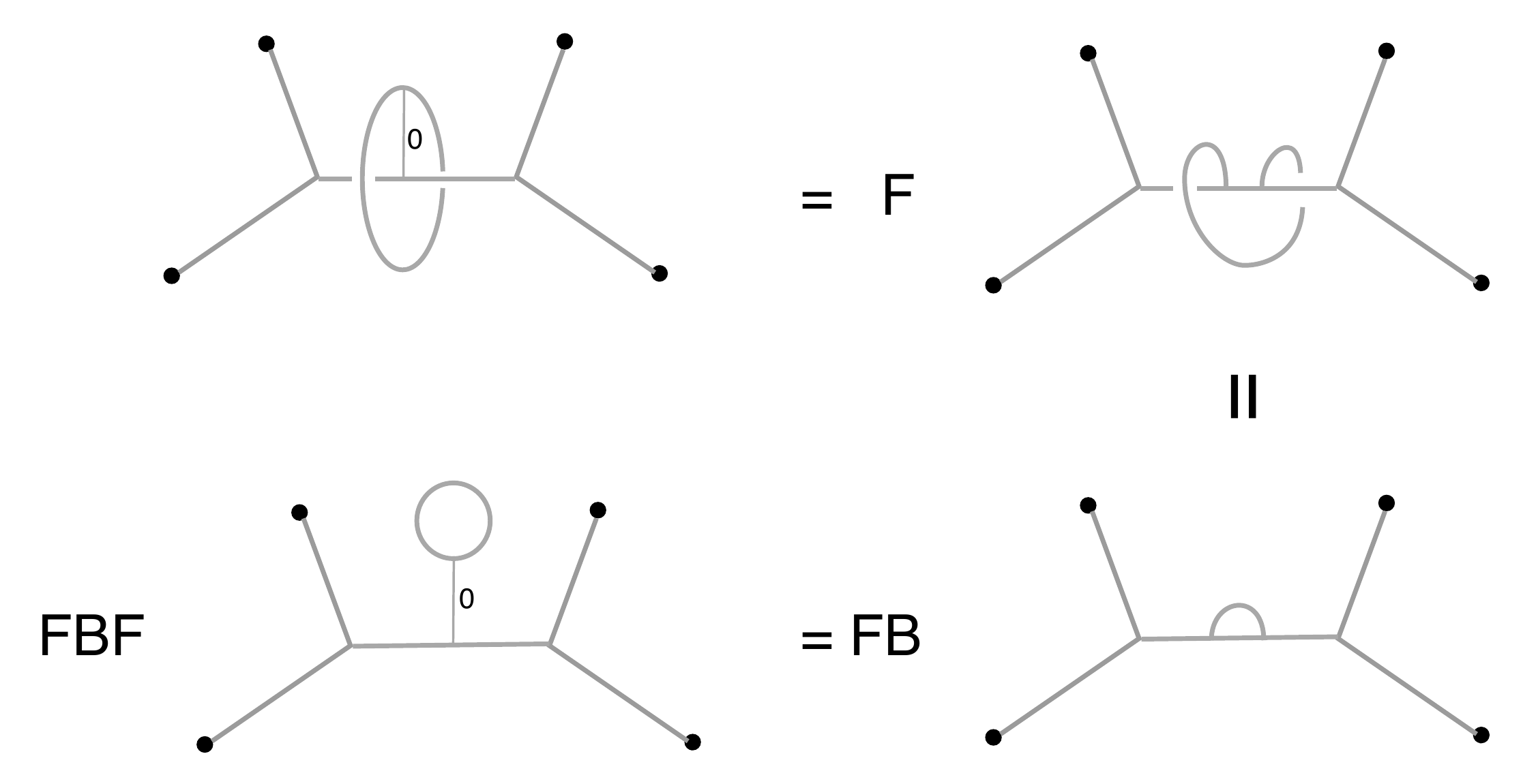}
\caption{\label{fig:4ptAdec} A graphical derivation of the sequence of fusion and braiding operations which realizes an A-cycle Verlinde line operator. }
\end{figure}

There is a simple strategy to compute $\hat O_\ell$. 
First one adds an extra line joining the new loop and the original graph, carrying a $(1,1)$ (identity) representation. 
Then one applies fusion and braiding transformations until the graph is simplified back to $\Gamma_0$, 
together contractible loops which can be eliminated. See Figure \ref{fig:4ptAdec} for the simplest example.
It is easy to see that this definition essentially reproduces the original definition of the Verlinde line defect.
Notice that $\hat O_\ell$ maps conformal blocks to conformal blocks with the same labels on external edges. 


\begin{figure}
\center \includegraphics[height=3cm]{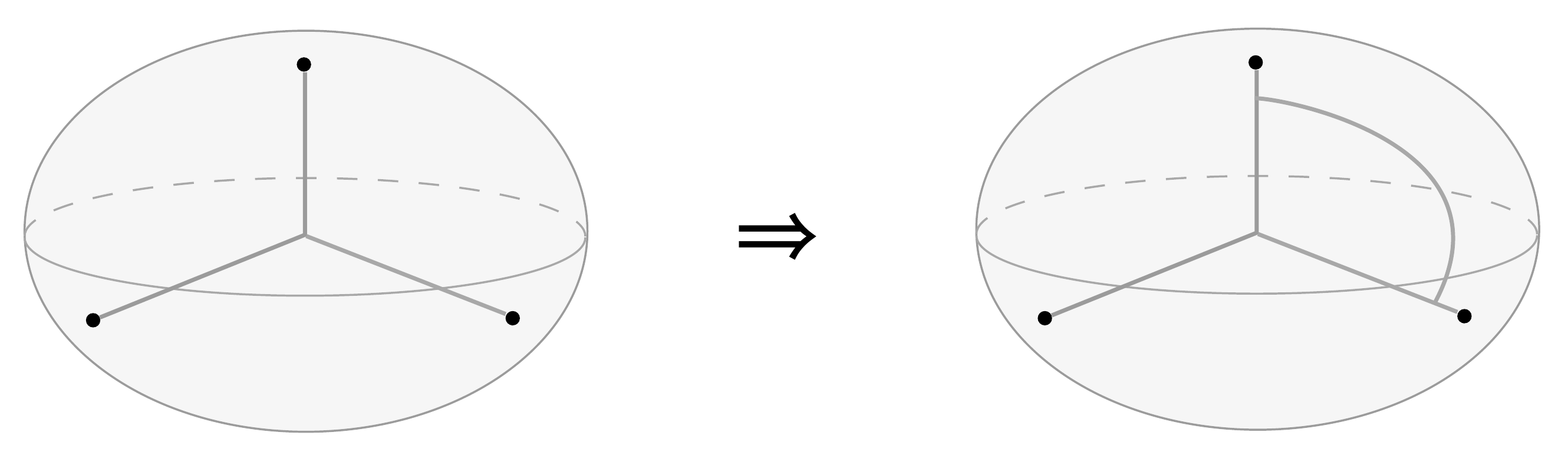}
\caption{\label{fig:3ptopen} The pictorial representation of the action of an open Verlinde line operator on a three point conformal block. 
We omit for clarity the Liouville momentum labels on the edges}
\end{figure}

The definition of open line operators follows along similar lines. The action of an open line operators is represented by adding a rung a the 
trivalent graph $\Gamma$ (without changing the geometry of the ambient 3-manifold!), which joins two external lines, say $e_1$ and $e_2$, 
and runs just below the boundary along the open path which labels the line operator. See Figure \ref{fig:3ptopen} for an example. 
The new rung splits each of the two external edges in two parts. On the part connected to the original $\Gamma$
we keep the original external momenta of $\Gamma$. On the external part, we allow generic shifts $s_1$, $s_2$ and $\tilde s_1$, $\tilde s_2$ 
of the external momenta. 

Thus the open line operator $\ell_{1,2}$ will be labelled by the choice of open path  and the shifts of external momenta. 
If we denote the new decorated graph as $\ell_{1,2} \circ \Gamma$, we can thus define 
\begin{equation}
\hat O_{\ell_{1,2}} \equiv \hat O_{\ell_{1,2} \circ \Gamma_0}
\end{equation}
It is easy to see it satisfies the relation:
\begin{equation}
|\ell_{1,2} \circ \Gamma;a_i\rangle_{\Gamma_0} = \hat O_{\ell_{1,2}} |\Gamma;a_i\rangle_{\Gamma_0}
\end{equation}
and that open and/or closed line operators compose nicely. 

It is important to notice that the operator  $\hat O_{\ell_{1,2}}$ include overall factors $p_1^{s_1}$, $p_2^{s_2}$, etc.
which shift appropriately the external momenta. As traditional spaces of conformal blocks are usually defined for fixed 
external momenta, the open Verlinde operators maps distinct traditional spaces into each other. 
Operators labelled open lines which do not intersect when projected on the Riemann surface will commute. 

\begin{figure}
\center \includegraphics[height=2cm]{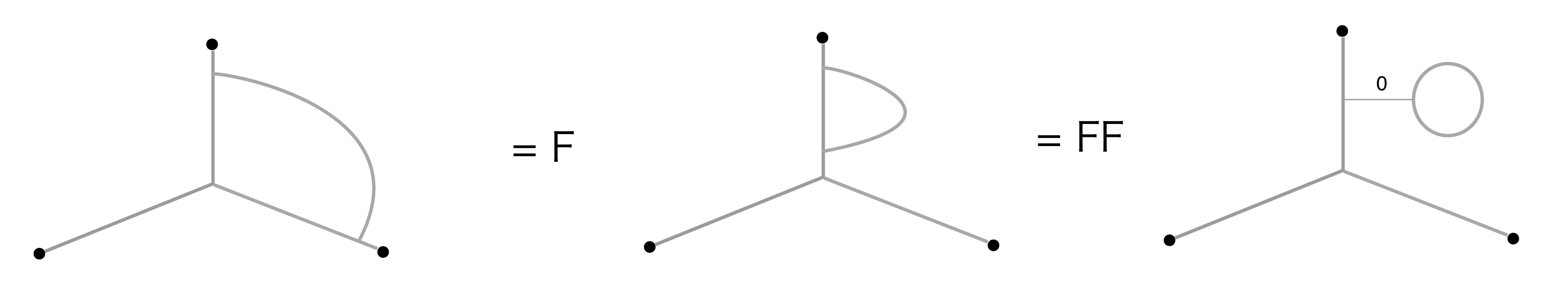}
\caption{\label{fig:3ptopendec} A graphical derivation of the sequence of fusion operations which realizes an open Verlinde line operator.}
\end{figure}

\section{Concrete calculations} \label{sec:for}

A conformal block is built up from a pair of pants decomposition of the Riemann surface. 
The concrete construction of a generic conformal block involves several choices, 
such as a choice of local coordinate at each of the punctures of the three-punctured spheres one is gluing up together. 
Most local changes in these conventions are essentially immaterial, as they can be absorbed in a re-definition of the 
local coordinates on Teichm\"uller space. Global changes, though, do matter. A rotation of $2 \pi$ in the local coordinate 
at an external puncture of conformal dimension $\Delta_a$, for example, will rotate the conformal block by a phase $e^{2 \pi i \Delta}$.

These facts can be expressed intrinsically as the statement that the conformal block is the section of some line bundle 
on Teichm\"uller space. Alternatively, one can pick some specific reference choices of local data, and keep track carefully of 
whenever these are changed. In either case, a useful way to keep track of the direction of the local coordinate system 
at a puncture which is being transported around is to use ribbon graphs, which keep track of how many times the local coordinate rotates around the origin. 

\begin{figure}
\center \includegraphics[height=3cm]{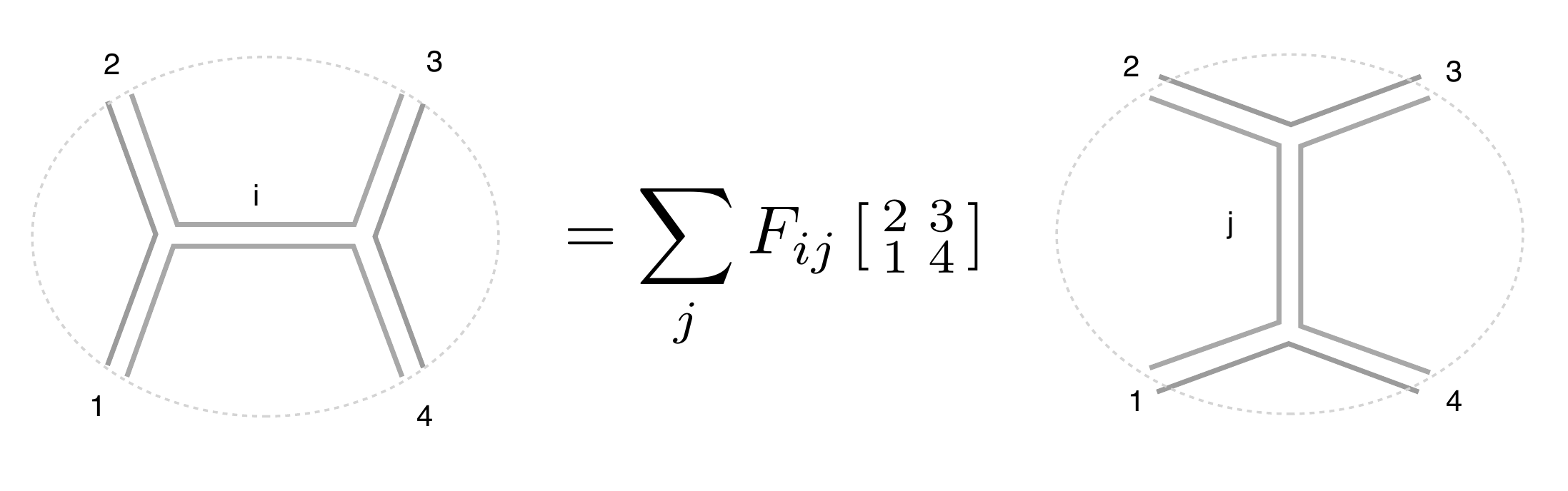} \includegraphics[height=3cm]{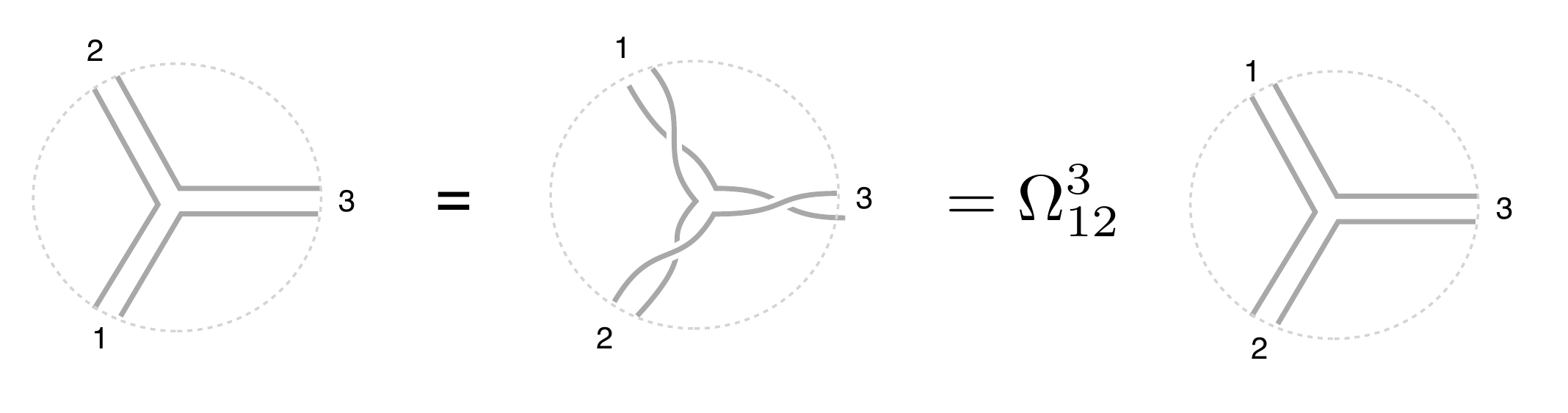}
\caption{\label{fig:ABmoves} The A and B moves, in a ribbon graph representation}
\end{figure}

In genus zero, there is a simple choice of reference coordinate systems: keep the punctures near the equator of the sphere, and 
point the positive real axis of the local coordinate always in the same direction along the equator. Then all ribbons are flat in the plane
of the picture, and we can draw them simply as lines without ambiguity. The basic manipulations of conformal blocs are A-moves, or fusion, which relates the two different (``s'' and ``t'' channel) planar 
way to connect four lines, and B-moves, which exchange (braid) the position of two legs of a trinion. See Figure \ref{fig:ABmoves}. 

More precisely, the B-move is a combination of a half-twist of one leg (say with label $i$), and a half-twist of the other two legs of the trinion (say with labels $j$ and $k$) in the opposite direction. This means that the two branches of the conformal block attached to $j$ and $k$ exchange place while the branch attached to $i$ remains fixed, but neither of the three branches rotates. The conformal block is multiplied by a phase
 \begin{equation}
 \Omega_{jk}^{i,\pm} =e^{\pm \i \pi  \left( \Delta_i - \Delta_j - \Delta_k \right)}
 \end{equation}
It is also helpful to think about this in terms of an OPE: a conformal block with external punctures of dimensions $\Delta_j$ and $\Delta_k$ 
fused to a channel of dimension $\Delta_i$ will behave as $(z_j - z_k)^{\Delta_i - \Delta_j - \Delta_k}$. Thus the conformal block equals $\Omega_{jk}^{i,+}$ times a conformal block with $k$ rotated clockwise around $j$. 
Notice that 
 \begin{equation}
 \Omega_{jk}^{i,+}  \Omega_{ik}^{j,+}= e^{-2\i \pi  \Delta_k}
 \end{equation}  

In the following, we will denote generic Virasoro representations by a Liouville momentum $Q/2 + \i a$, so that 
\begin{equation}
\Delta_i = \frac{Q^2}{4} + a_i^2
\end{equation}
and $Q = b + b^{-1}$, while the central charge is $1+ 6 Q^2$. We will also define 
\begin{equation}
q = e^{\i \pi b^2} \qquad \qquad \tilde q = e^{\i \pi b^{-2}}.
\end{equation}

Degenerate representations are labeled as $(r,s)$, with Liouville momentum $a_{(r,s)} = \i r b/2 + \i s b^{-1}/2$, and corresponding dimension 
\begin{equation}\Delta_{r,s} = \left( (1- r) b/2 +(1- s) b^{-1}/2\right) \left( (1+ r) b/2 +(1+ s) b^{-1}/2\right).\end{equation} 
The identity operator has label $(1,1)$. Unless otherwise specified, we will focus on the degenerate field $(2,1)$, indicated as a thin line in the figures. 
The basic building blocks we will use are the A and B moves involving a single $(2,1)$ degenerate field. 

\begin{figure}
\center \includegraphics[height=6cm]{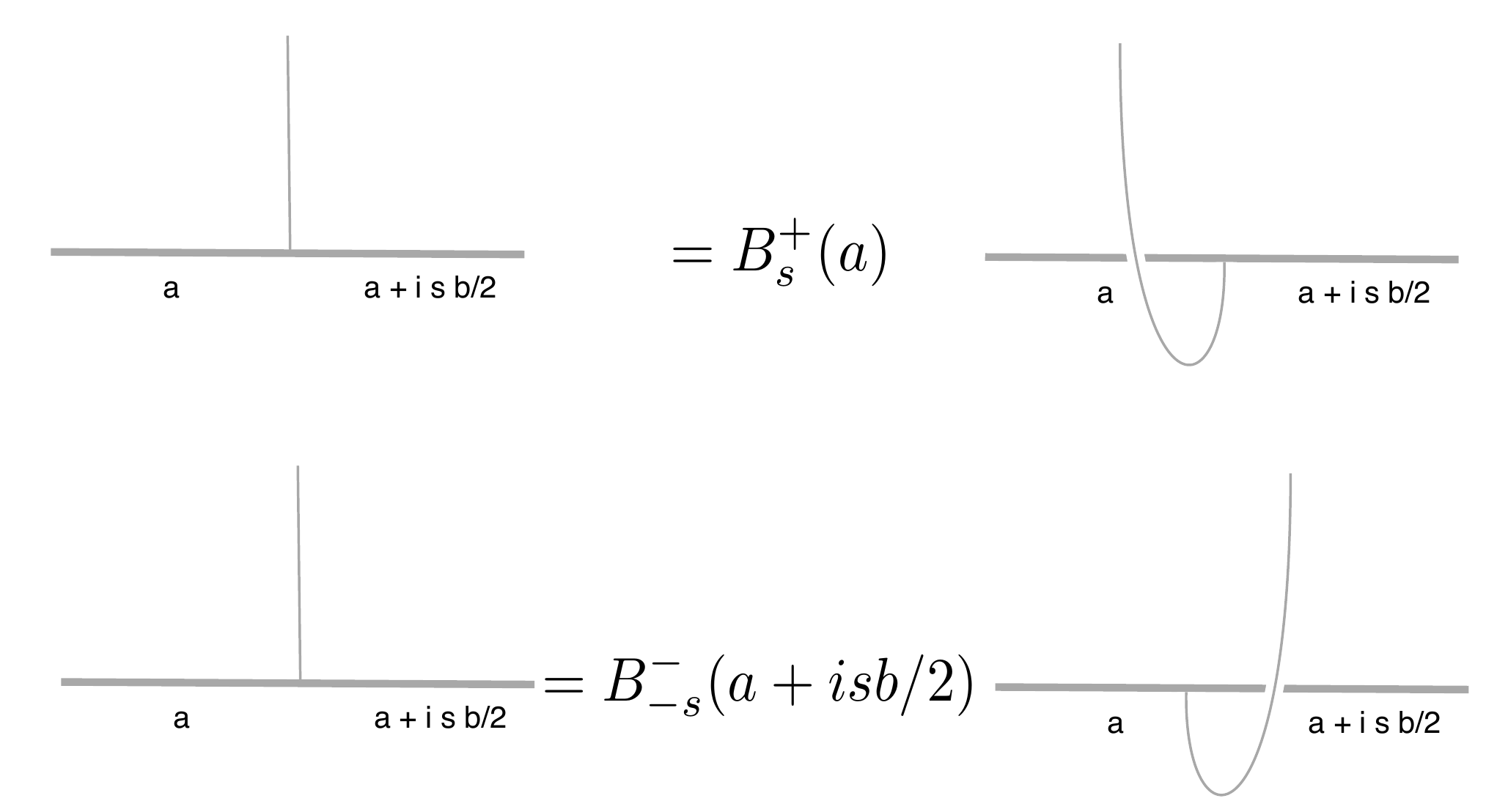}
\caption{\label{fig:Bdegen} Two examples of a B move on a degenerate leg}
\end{figure}

If we specialize the $k$ leg to be degenerate in 
$ \Omega_{jk}^{i,\pm}$, and impose the degenerate fusion relation $a_j = a$, $a_i = a + i s b/2$, $s = \pm 1$,  we get 
 \begin{equation}
B^\pm_s(a) = \Omega_{j,(2,1)}^{i,\pm} =e^{\pm \i \pi  \left( \i s b a + b^2/2 + 1/2 \right)} \equiv (-q)^{\pm \frac{1}{2}} \alpha^{\mp s} 
 \end{equation}
Remember that $q = e^{\i \pi b^2}$ and $\alpha = e^{\pi b a}$.
It is useful to visualize this phase as being accrued when a degenerate insertion is moved from one side to the other of a non-degenerate leg of the graph, as in figure \ref{fig:Bdegen}. 

\begin{figure}
\center \includegraphics[height=3cm]{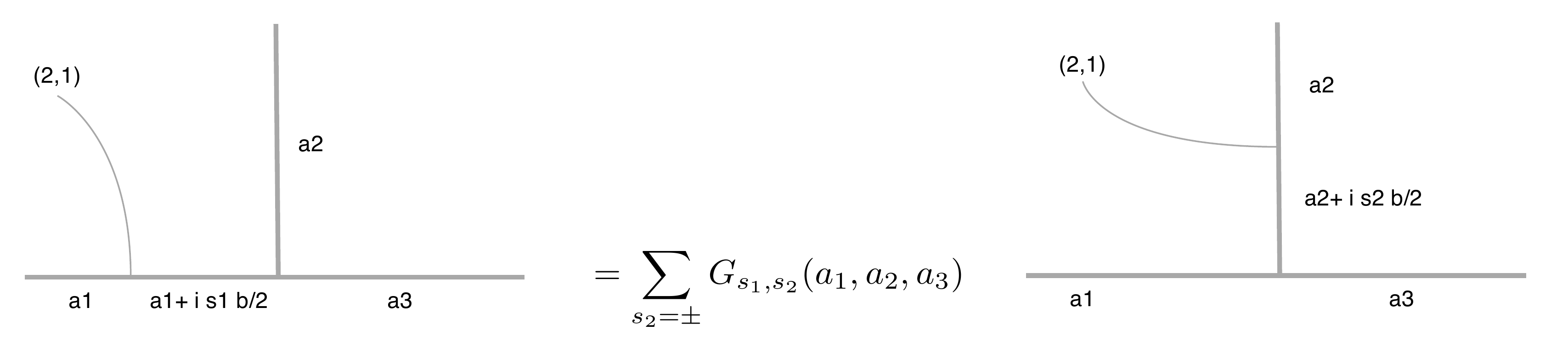}
\caption{\label{fig:Adegen} The A move on a degenerate leg}
\end{figure}

Next, consider a four-point conformal block on the sphere, with three generic legs of label $a_1$, $a_2$, $a_3$ and a single degenerate insertion. 
We want to consider the fusion matrix 
\begin{equation}
G_{s_1, s_2}(a_1,a_2,a_3) \equiv F_{a_1 + \i s_1 b/2,a_2 + \i s_2 b/2} \left[ \begin{smallmatrix} a_{(2,1)} & a_2 \cr a_1 & a_3\end{smallmatrix}\right] 
\end{equation}
which transports a degenerate insertion from the $a_1$ to the $a_2$ leg of a trivalent graph, as in figure \ref{fig:Adegen}.
With an appropriate choice of normalization for the three point function, the matrix is  
\begin{equation}
G_{s_1, s_2}(a_1,a_2,a_3) = \i s_2\frac{\cosh \pi b ( a_3 + s_1 a_2 - s_2 a_1) }{\sinh 2\pi b a_2} = \i s_2 \frac{\alpha_3 \alpha_2^{s_1} \alpha_1^{- s_2}+\alpha_3^{-1} \alpha_2^{-s_1} \alpha_1^{ s_2}}{\alpha_2^2- \alpha_2^{-2} }
\end{equation}

\begin{equation}
\sum_{s_2=\pm 1} G_{s_1, s_2}(a_1,a_2,a_3)G_{s_2, s_1'}(a_2,a_1,a_3) = - s'_1\frac{\alpha_2^{- s_1- s_1'}  - \alpha_2^{ s_1+s_1'}}{\alpha_2^2- \alpha_2^{-2} } = \delta_{s_1,s_1'}
\end{equation}

\begin{figure}
\center \includegraphics[height=8cm]{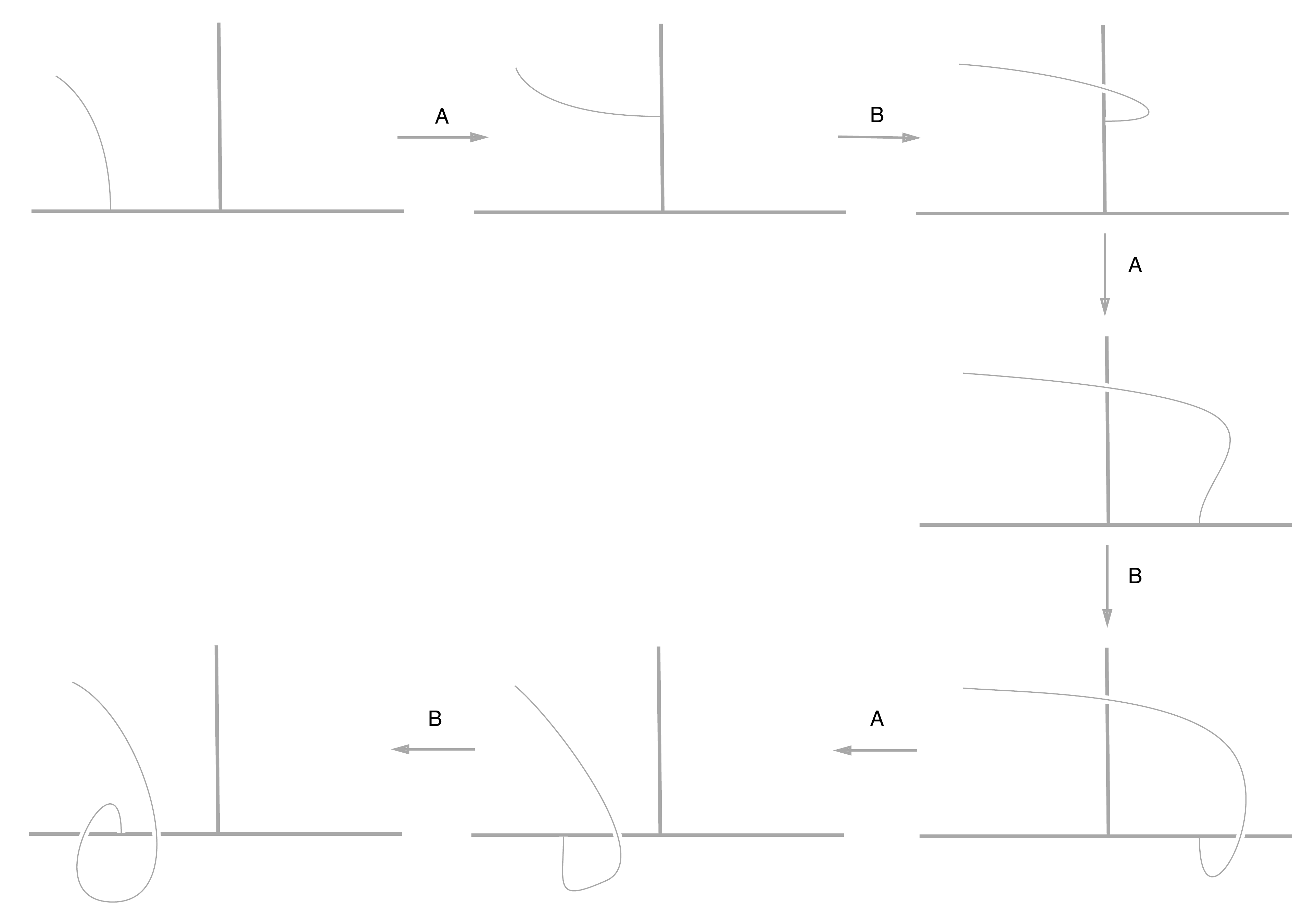}
\caption{\label{fig:Hexa} This sequence of A and B moves brings the conformal block back to itself up to a full twist of the degenerate leg}
\end{figure}

A second check is an hexagon relation, where we alternate three A and three B moves in order to transport the degenerate field around a closed loop: 
 \begin{align}
\sum_{s_2,s_3} G_{s1,s2} ( a_1,a_2,a_3) &B^\pm_{s_2}(a_2) G_{s2,s3} ( a_2,a_3,a_1) B^\pm_{s_3}(a_3)G_{s3,s1'} ( a_3,a_1,a_2)  B^\pm_{s_3}(a_3) \cr &= - e^{ \pm \frac{3}{2} i \pi b^2} \delta_{s_1, s_1'}. \end{align}
 The phase factor originates from the fact that the local coordinate 
around the degenerate field is rotated by an angle of $2 \pi$ in the process, as in figure \ref{fig:Hexa}. 

\begin{figure}
\center \includegraphics[height=3cm]{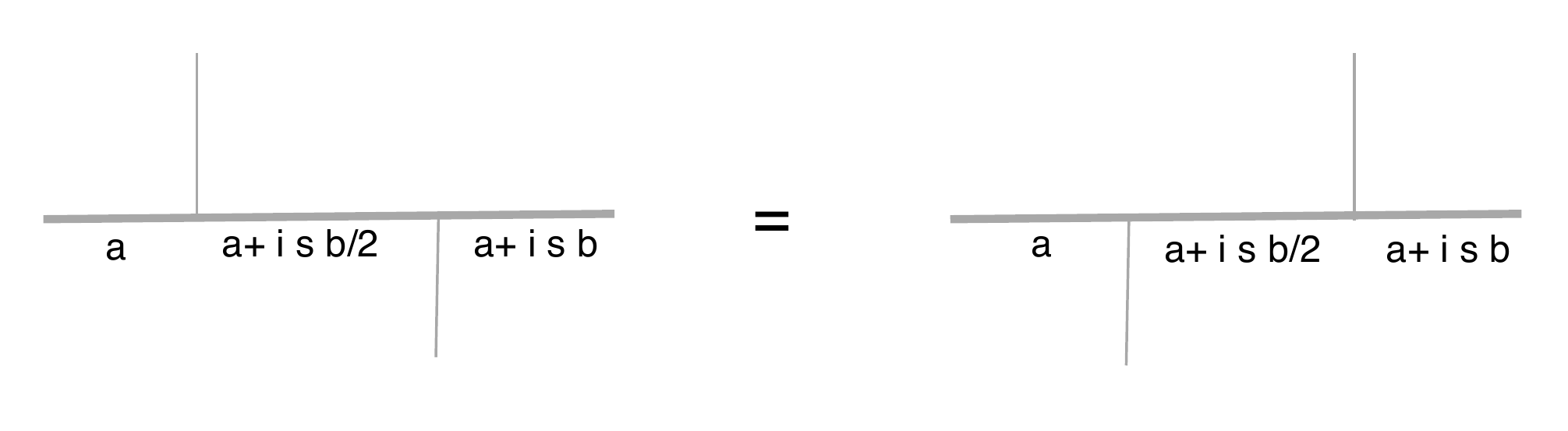}
\caption{\label{fig:com} Two degenerate insertions on opposite sides commute if they shift the Liouville momentum in the same direction}
\end{figure}

We will need a few specializations of this formula as well. First, consider a situation where the third puncture is degenerate, $a_3 = \i b + \i b^{-1}/2$,
 and the other two satisfy the degenerate fusion so that $a_1= a$ and $a_2 = a + \i s_1 b/2 + \i s b/2$. We are interested in the situation where $s = s_1$, i.e. where the two degenerate insertions shift the momentum $a$ twice in the same direction. 
 Then after the $A$ move we must have $s_2 = -s_1$ as well, and the fusion matrix reduces to $1$: two degenerate insertions which and on opposite sides of a line, and shift the momentum in the same direction commute. See figure \ref{fig:com}.

\begin{figure}
\center \includegraphics[height=4.5cm]{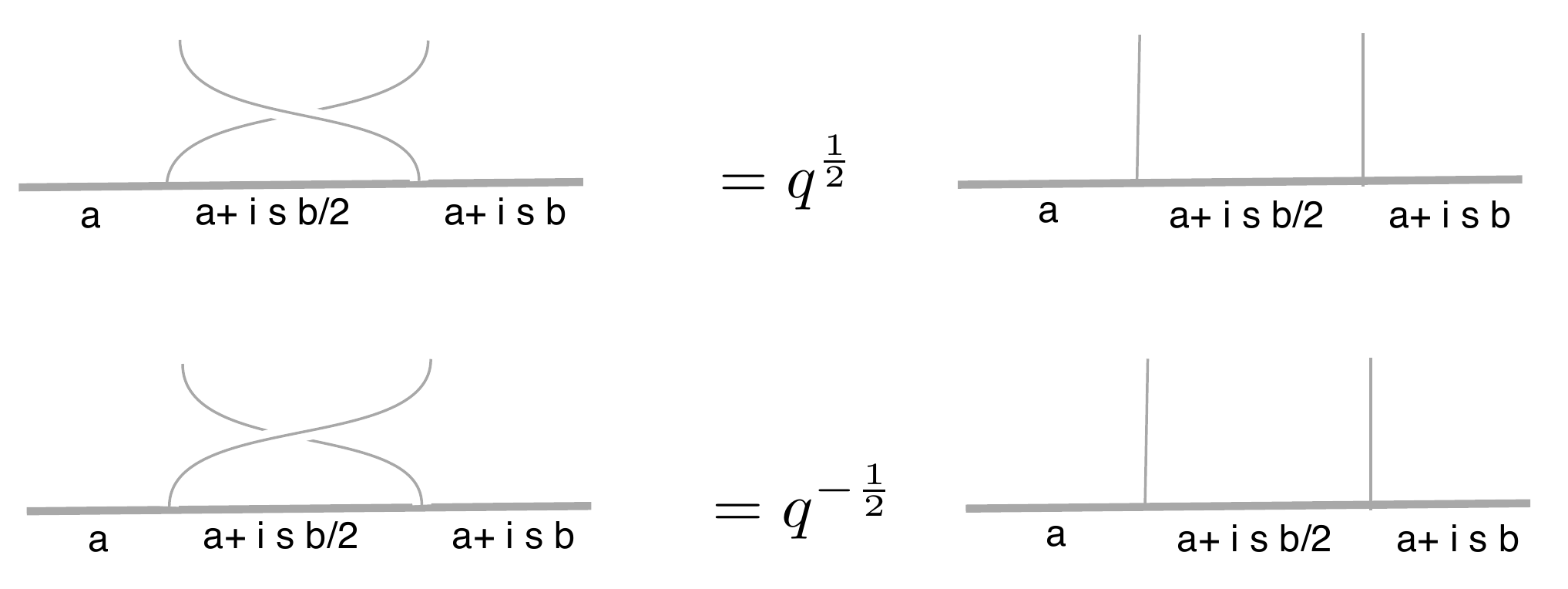}
\caption{\label{fig:com2} Two degenerate insertions on the same sides Weyl-commute if they shift the Liouville momentum in the same direction. The sign depends on the direction of the braiding}
\end{figure}

Two degenerate insertions on the same side of an edge, which shift the Liouville momentum in the same direction, can be commuted across each other with the help of two braiding moves, to move one puncture on the opposite side of the edge and back. 
The result is that two such insertions Weyl commute, as in figure \ref{fig:com2}.

This can also be understood as follows: the two degenerate insertions can be fused by an A move, braided and unfused by another A-move. Although the OPE of two $(2,1)$ fields would usually contain both the identity and the $(3,1)$ field, 
the shift in the Liouville momenta imply that only the $(3,1)$ channel can contribute. The OPE in the $(3,1)$ channel behaves as $(z_1 - z_2)^{- \frac{b^2}{2}}$, and thus the braiding produces factors of $q^{\pm \frac{1}{2}}$. 

\begin{figure}
\center \includegraphics[height=4.5cm]{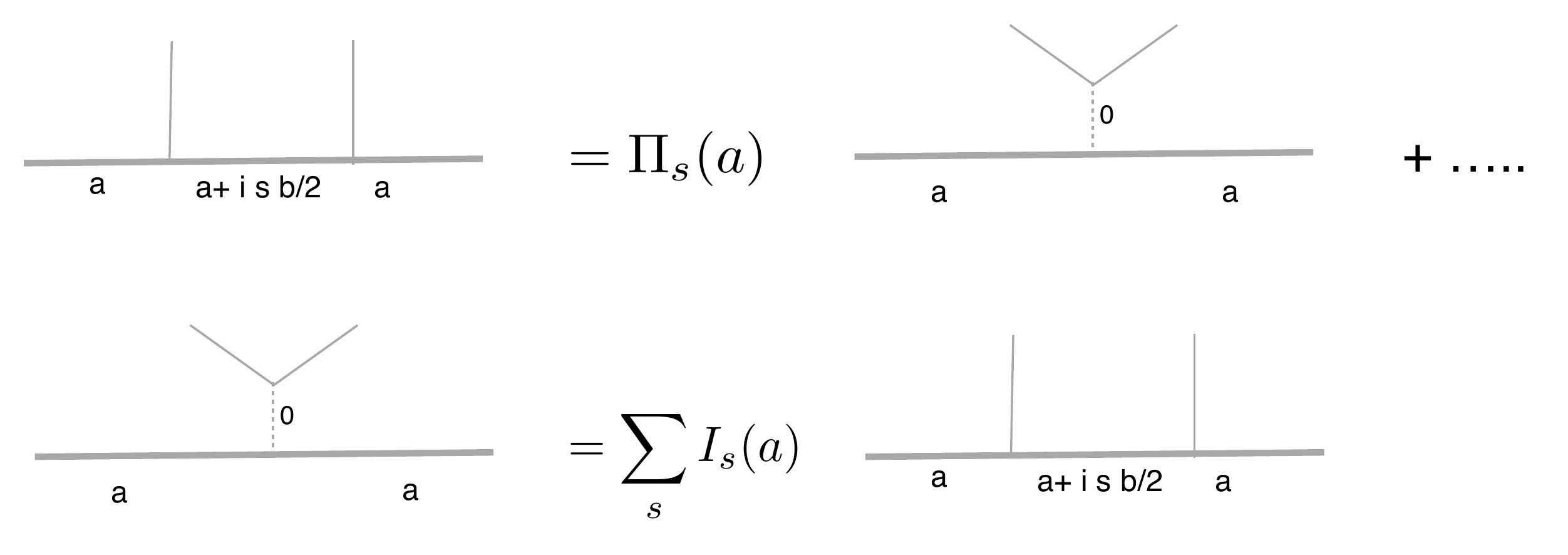}
\caption{\label{fig:ide} Fusion of degenerate fields to the identity and back. The ellipsis indicates an intermediate $(3,1)$ channel}
\end{figure}

If we take the second leg to be degenerate instead, we get the projection operator which can be used to fuse two nearby degenerate fields to the identity channel:
we set $a_2 = \i b + \i b^{-1}/2$, $s_2=-1$ and $a_3 = a_1=a$, $s_1=s$ to get
\begin{equation}
\Pi_{s}(a) = - s \frac{\sinh \pi b ( 2a + \i s b) }{\sinh 2\pi \i b^2}   = - s \frac{\alpha^2 q^s - \alpha^{-2} q^{-s}}{q^{2} - q^{-2}}
\end{equation} 
Viceversa, if we take the first leg to be degenerate, we get the map which splits an identity into two degenerate insertions: we set $a_1 = \i b + \i b^{-1}/2$, 
$s_1 = -1$, $a_3 = a_2=a$ and $s_2 = s$ to get 
\begin{equation}
I_{s}(a) = -s \frac{\sinh \i \pi b^2}{\sinh 2\pi b a}=  -s \frac{q - q^{-1}}{\alpha^2 - \alpha^{-2}}
\end{equation} 
See figure \ref{fig:ide}

It is also useful to derive the skein relation for degenerate lines. We can start from the second relation in figure \ref{fig:ide}, specialized to $a=a_{(2,1)} = \i b + \i b^{-1}/2$, 
so that we expand the identity t-channel conformal block $f_t$ in s-channel blocks.

\begin{figure}
\center \includegraphics[height=2.5cm]{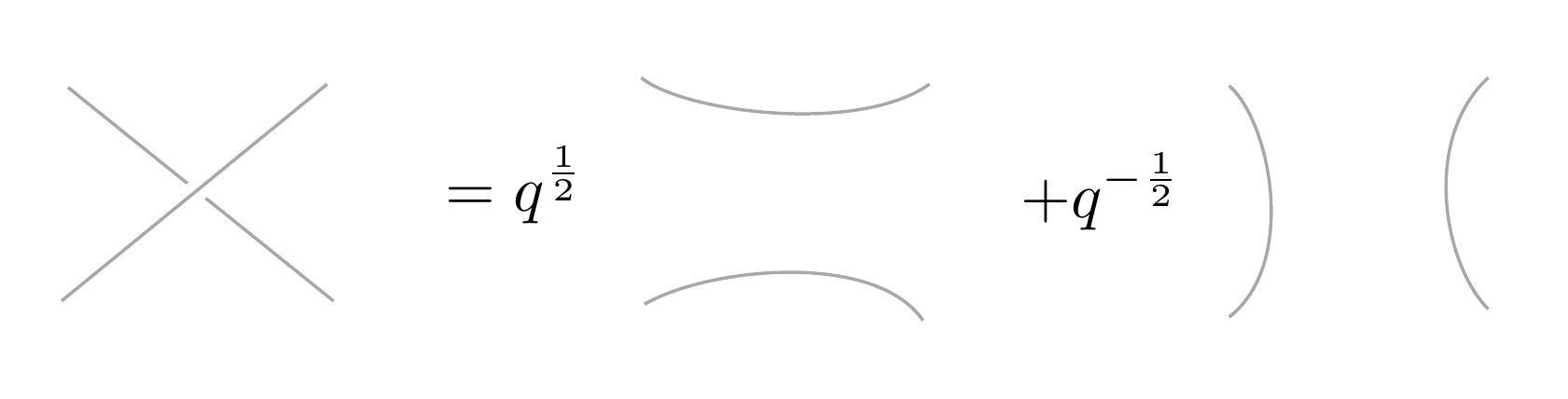}
\caption{\label{fig:skein} Skein relation for $(2,1)$ degenerate edges}
\end{figure}

\begin{figure}
\center \includegraphics[height=1cm]{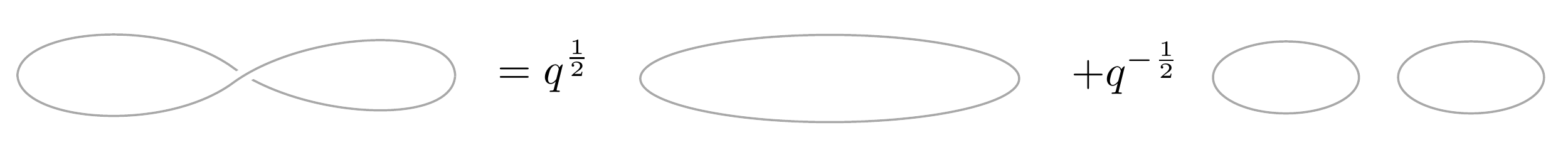}
\caption{\label{fig:eight} A closed contractible loop evaluates to a quantum dimension which is determined from the skein relation applied to a figure eight}
\end{figure}

The coefficient $I_{-1}(a_{(2,1)})$ of the identity s-channel conformal block $f_s$ is $-(q + q^{-1})^{-1}$. We can braid the two degenerate legs on the left to get the expansion of an identity u-channel block $f^\pm_u$, and take a linear combination of the two formulae 
to project away the $(3,1)$ s-channel block.
The result is 
\begin{equation}
f^\pm_u - q^{\pm 1/2} f_t = (q + q^{-1})^{-1} q^{\mp 3/2} \left( 1 + q^{\pm 2} \right) f_s= q^{\mp 1/2} f_s
\end{equation}
which is the expected skein relation, depicted in figure \ref{fig:skein}.
If we apply the skein relation to a figure eight closed loop, as in figure \ref{fig:eight}, we get a constraint on the quantum dimension $d_{(2,1)}$:  a contractible loop evaluates to
\begin{equation}
d_{(2,1)} = - q - q^{-1}
\end{equation} 

This is all the information we need to compute open and closed Verlinde line operators. 

\section{Examples}\label{sec:ex}
\begin{figure}
\center \includegraphics[height=1.5cm]{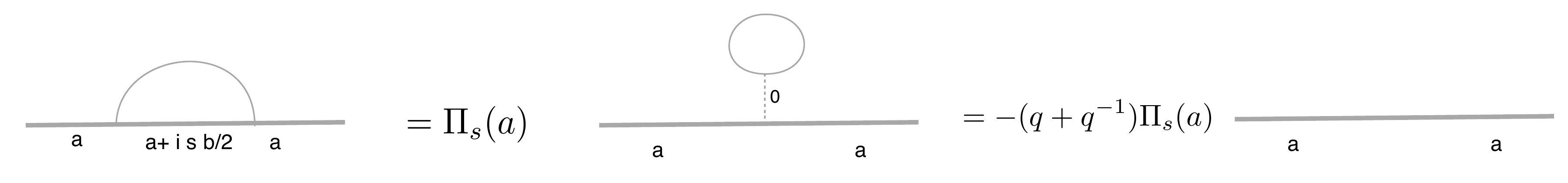}
\caption{\label{fig:bub} An almost trivial ``bubble'' decoration}
\end{figure}

\begin{figure}
\center \includegraphics[height=1.5cm]{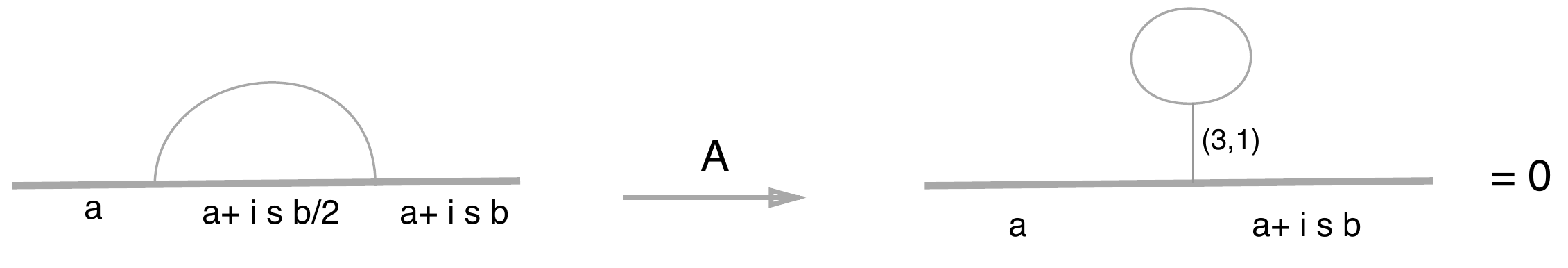}
\caption{\label{fig:empty} An ``bubble'' decoration which vanishes due to our axioms}
\end{figure}

It is useful to start from a somewhat trivial example of decorated graph $\Gamma$: 
a single extra degenerate line which starts and ends on the same edge of a graph $\Gamma_0$, as in figure \ref{fig:bub}.
We may simply fuse the two insertions to produce a tadpole graph, which by our axioms projects the intermediate channel to the identity.
We can thus remove the identity line, and contract away the closed degenerate loop to remove the degenerate line. The final result is a conformal block without decorations, multiplied by the ``bubble factor''
\begin{equation}
A_s[a]|a\rangle_{\Gamma_0}  = -(q + q^{-1})\Pi_{s}(a)|a\rangle_{\Gamma_0}  = \frac{\alpha^{2s} q - \alpha^{-2 s} q^{-1}}{q - q^{-1}}|a\rangle_{\Gamma_0} 
\end{equation}
Notice that a similar bubble diagram with different Liouville momenta in the external legs, as in figure \ref{fig:empty}, will always vanish because of our axioms:
the degenerate fusion requires the intermediate channel attached to the closed loop differ from the identity, and it is thus projected to zero. 


\begin{figure}
\center \includegraphics[height=1.5cm]{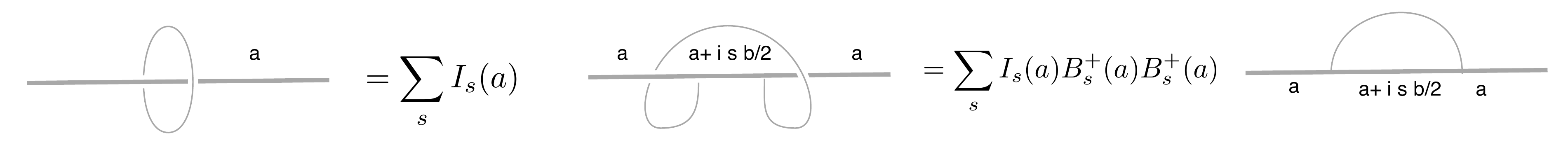}
\caption{\label{fig:wilson} The reduction of a basic closed Verlined line defect to a bubble decoration}
\end{figure}

The most basic example of closed Verlinde line operator is associated to an A-type loop, a closed degenerate loop winding an edge of the graph $\Gamma_0$, as in figure \ref{fig:wilson}.
It can be evaluated to 
\begin{equation}
\sum_s \Pi_s(a) B_s^+(a)B_s^+(a) A_s[a]|a\rangle_{\Gamma_0} = \left( \alpha^2 + \alpha^{-2} \right)|a\rangle_{\Gamma_0} 
\end{equation}
It is also possible to compute the multiplicative coefficient for a multiple-wound loop:
\begin{equation}
\sum_s \Pi_s(a) \left[B_s^+(a)\right]^{2n} \frac{\alpha^{2s} q - \alpha^{-2 s} q^{-1}}{q - q^{-1}} = q^{n-1} \frac{\alpha^{2n+2} - \alpha^{-2n-2}}{\alpha^2 - \alpha^{-2}} - q^{n+1} \frac{\alpha^{2n-2} - \alpha^{-2n+2}}{\alpha^2 - \alpha^{-2}}
\end{equation}

\begin{figure}
\center \includegraphics[height=3cm]{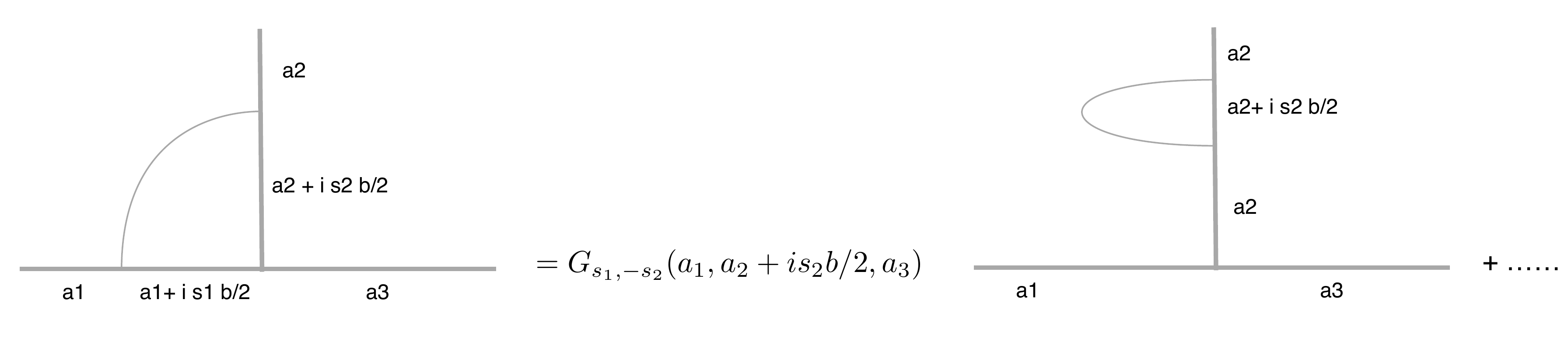}
\caption{\label{fig:openwil} The computation of the triangle diagram $T_{s_1,s_2}$ used in the definition of the open Verlinde line defect ${\cal O}^{-s_1,-s_2}_{1,2}$. The ellipsis indicates a term with Liouville momentum $a_2 + \i s_2 b$ in the fusion channel, which vanishes as in figure \ref{fig:empty}}
\end{figure}

Next, we can look at examples involving a single trinion in $\Gamma_0$. 
It can be computed from a triangle diagram $T_{s_1,s_2}$ with a degenerate line connecting consecutive edges of a trinion $T_0$, as in figure \ref{fig:openwil}.
If we write 
\begin{equation}
|T_{s_1,s_2};a_1,a_2,a_3\rangle = T_{s_1,s_2}[a_1,a_2,a_3] |a_1,a_2,a_3\rangle
\end{equation}
then we can compute 
\begin{equation}
T_{s_1,s_2}[a_1,a_2,a_3] = -\i \frac{\alpha_3 \alpha_2^{s_1} \alpha_1^{s_2} q^{\frac{s_1 s_2}{2}} + \alpha_3^{-1} \alpha_2^{-s_1} \alpha_1^{-s_2} q^{\frac{-s_1 s_2}{2}} }{q - q^{-1}} \end{equation}

We can do some checks of this expression, and our formalism, by specializing some of the external legs to 
degenerate values and imposing degenerate fusion. For example, if we take the second leg to be degenerate of type $(1,1)$, i.e. the identity, so that we have $\alpha_2 = \i q^{1/2}$, $s_2 = 1$, $s_1 = s$ and $a_1 = a_3 =a$, the triangle reduces to the the bubble factor. If we take the second leg to be degenerate of type $(2,1)$, $\alpha_2 = \i q$, $s_2 = -1$, $\alpha_3 = q^{s_1/2} \alpha_1$,
so that the intermediate momentum on the second leg is that of the identity, we get $1$. 

If we take the second leg to be degenerate of type $(3,1)$, $\alpha_2 = \i q^{3/2}$, $s_2 = -1$, $\alpha_3 = q^{s_1/2+s'_1/2} \alpha_1$, so that the intermediate momentum on the second leg is of type $(2,1)$, we get a $(3,1)$ external leg which 
splits into two $(2,1)$ legs and the corresponding factor is $q + q^{-1}$ if $s_1 = s_2$, $1$ otherwise. 
We can call the latter special triangle coefficient ``splitting factor'' $S_{s_1,s'_1}[a_1]$.
This is a very useful piece of information to study graphs which include a $(3,1)$ leg. 

For example, consider a trinion with two consecutive $(2,1)$ rungs inserted on it, of shifts $s_1$, $s_2$ and
$s'_1$, $s'_2$ respectively. We can reduce it to the basic undecorated trinion multiplied by 
\begin{equation}
 T_{s_1,s_2}[a_1,a_2,a_3]T_{s'_1,s'_2}[a_1+ \i s_1 b/2,a_2+ \i s_2 b/2,a_3].
\end{equation}
On the other hand, we can add a trivial $(1,1)$ rung between the two parallel $(2,1)$ lines and fuse it into 
a $(3,1)$ line or an $(1,1)$ line. The former can be simplified to a simple triangle diagram with a $(3,1)$ 
rung with the help of the splitting factors. The latter can be simplified with two bubble factors. 
We expect thus a relation 
\begin{align}
 T_{s_1,s_2}&[a_1,a_2,a_3]T_{s'_1,s'_2}[a_1+ \i s_1 b/2,a_2+ \i s_2 b/2,a_3] = \cr &= I_{1}(a_{(2,1)})S_{s_1,s'_1}[a_1]S_{s_2,s'_2}[a_2]T^{(3,1)}_{s_1+s'_1,s_2+s'_2}[a_1,a_2,a_3] +  \cr &+I_{0}(a_{(2,1)})\delta_{s_1 +s'_2,0}\delta_{s_2 + s'_2,0} A_{s_1}(a_1)A_{s_2}(a_2)
\end{align} 

It is easy to check that these $16$ relations are compatible with each other, and define uniquely the $9$ operators $T^{(3,1)}_{s_1,s_2}[a_1,a_2,a_3] $. 
This procedure can be iterated to derive systematically triangles $T^{(k,1)}_{s_1,s_2}[a_1,a_2,a_3]$ with the extra rungs 
labelled by higher degenerate representations. 

We are ready to write down the open Verlinde line operator ${\cal O}^{s_1,s_2}_{1,2}$ acting on the $1$ and $2$ legs of the trinion. 
\begin{align}
{\cal O}^{s_1,s_2}_{1,2} &|a_1,a_2,a_3\rangle =  |T_{-s_1,-s_2};a_1+ \i s_1 b/2,a_2+ \i s_2 b/2,a_3\rangle = \cr 
&=  -\i \frac{\hat \alpha_3 \hat \alpha_2^{-s_1} \hat \alpha_1^{-s_2} q^{\frac{s_1 s_2}{2}} + \hat \alpha_3^{-1} \hat \alpha_2^{s_1} \hat \alpha_1^{s_2} q^{\frac{-s_1 s_2}{2}} }{q - q^{-1}} \hat p_1^{s_1} \hat p_2^{s_2} |a_1,a_2,a_3\rangle
\end{align}

An important and non-trivial check is the observation that the operators ${\cal O}^{s_1,s_2}_{1,2}$,  ${\cal O}^{s_2,s_3}_{2,3}$ and  ${\cal O}^{s_3,s_1}_{3,1}$ for given $s_i$ acting on a three-punctured sphere commute, 
as they should, as they end on opposite sides of the same line, and shift the Liouville momentum in the same direction. 
For each choice of three $s_i$ we can find an alternative normalization of the three-point conformal block such that the ${\cal O}^{s_i,s_j}_{i,j}$ reduce to $p_i^{s_i} p_j^{s_j}$. 

\begin{figure}
\center \includegraphics[height=3cm]{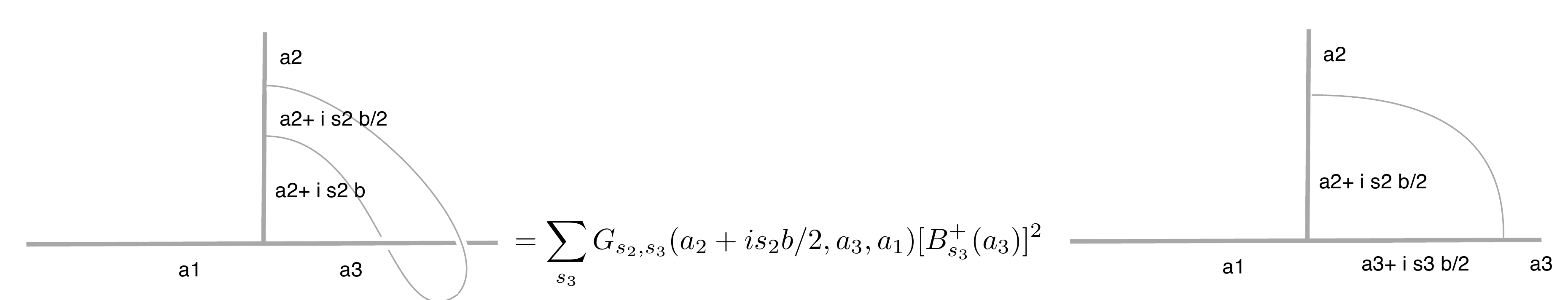}
\caption{\label{fig:loopy} The computation of the decorated trinion $L_{s_2}$ used in the definition of an open Verlinde line defect with both ends on the same external leg.}
\end{figure}

Later, we will also need to consider an open Verlinde line operator we can associate to a trinion, which starts and ends on the same leg, and shifts the corresponding momentum twice in the same direction, is based on figure \ref{fig:loopy}.
We can compute the appropriate coefficient here:
\begin{align}
L_{s_2}[a_1,a_2,a_3]=&\sum_{s_3}s_3 \frac{\alpha_1 \alpha_3^{s_2} \alpha_2^{- s_3} q^{-1/2}+\alpha_1^{-1} \alpha_3^{-s_2} \alpha_2^{ s_3} q^{1/2}}{\alpha_3^2- \alpha_3^{-2} } \cdot \cr &(-q) \alpha_3^{- 2 s_3} \frac{\alpha_1 \alpha_3^{s_2} \alpha_2^{s_3} q^{\frac{s_2 s_3}{2}} + \alpha_1^{-1} \alpha_3^{-s_2} \alpha_2^{-s_3} q^{\frac{-s_2 s_3}{2}} }{q - q^{-1}} 
\end{align}

\subsection{Four-punctured sphere}
Next, we can consider the action of open Verlinde line operators on a four-point sphere s-channel conformal block, 
as in the top left corner of Figure \ref{fig:ABmoves}. The open Verlinde line operators joining directly the $1$ and $2$ edges, 
or the $3$ and $4$ edges, $O^{s_1,s_2}_{1,2}$ and $O^{s_3,s_4}_{3,4}$, are obviously the same as we found for the trinion:
\begin{align}
{\cal O}^{s_1,s_2}_{1,2} |a_i;a\rangle &=  -\i \frac{\hat \alpha \hat \alpha_2^{-s_1} \hat \alpha_1^{-s_2} q^{\frac{s_1 s_2}{2}} + \hat \alpha^{-1} \hat \alpha_2^{s_1} \hat \alpha_1^{s_2} q^{\frac{-s_1 s_2}{2}} }{q - q^{-1}} \hat p_1^{s_1} \hat p_2^{s_2} |a_i;a\rangle \cr
{\cal O}^{s_3,s_4}_{3,4} |a_i;a\rangle &=  -\i \frac{\hat \alpha \hat \alpha_4^{-s_3} \hat \alpha_3^{-s_4} q^{\frac{s_3 s_4}{2}} + \hat \alpha^{-1} \hat \alpha_4^{s_3} \hat \alpha_3^{s_4} q^{\frac{-s_3 s_4}{2}} }{q - q^{-1}} \hat p_3^{s_3} \hat p_4^{s_4} |a_i;a\rangle
\end{align}

\begin{figure}
\center \includegraphics[height=3cm]{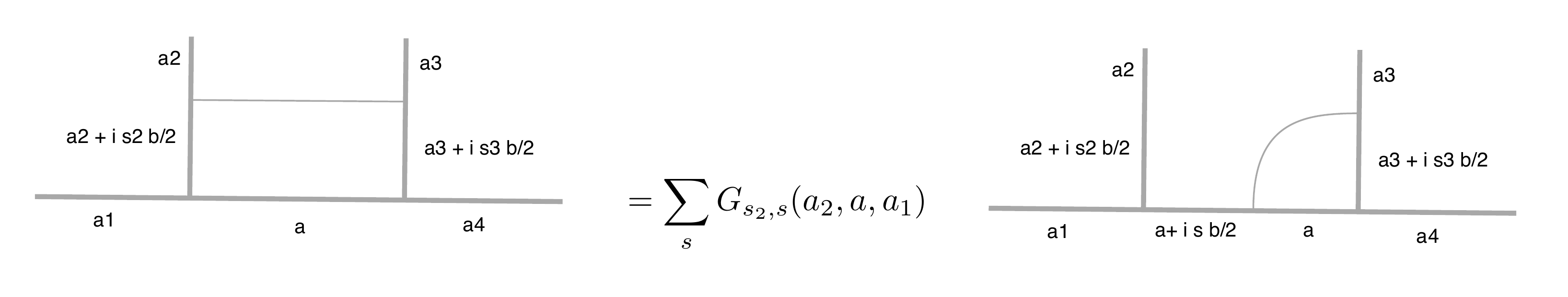}
\caption{\label{fig:4pt23} The computation of the coefficients for a rectangular graph conformal block $R^s_{s_2,s_3}[a_1,a_2,a_3,a_4,a]$ }
\end{figure}

There are two natural operators which join directly the punctures $2$ and $3$ and $1$ and $4$, which we can denote as 
$O^{s_2,s_3}_{2,3}$ and $O^{s_4,s_1}_{4,1}$. 
The operator $O^{s_2,s_3}_{2,3}$ is computed from the coefficients for a rectangular graph in figure \ref{fig:4pt23} as
\begin{align}
O^{s_2,s_3}_{2,3} &|a_i,a \rangle = \sum_{s=\pm 1} \hat p^s R^s_{-s_2,-s_3}[\hat a_1,\hat a_2,\hat a_3,\hat a_4,\hat a] \hat p_2^{s_2}\hat p_3^{s_3}|a_i;a \rangle
\end{align}
where
\begin{align}
R^s_{s_2,s_3}&[a_1,a_2,a_3,a_4,a] =G_{s_2,s}(a_2,a,a_1) T_{-s,s_3}[a+ \i s b/2,a_3,a_4]  \cr 
&= s \frac{\alpha_1 \alpha^{s_2} \alpha_2^{- s}+\alpha_1^{-1} \alpha^{-s_2} \alpha_2^{ s}}{\alpha^2- \alpha^{-2} }\frac{\alpha_4 \alpha_3^{-s} \alpha^{s_3} + \alpha_4^{-1} \alpha_3^{s} \alpha^{-s_3}}{q - q^{-1}}
\end{align}
Notice the position of the $\hat p$ operators in the expression for $O^{s_2,s_3}_{2,3}$, and remember that 
\begin{equation}
\hat p \hat a = (\hat a - \i b/2) \hat p \qquad \qquad \hat p_i \hat a_i = (\hat a_i - \i b/2) \hat p
\end{equation}

Similarly, 
\begin{align}
O^{s_4,s_1}_{4,1} &|a_i,a \rangle = \sum_{s=\pm 1} \hat p^s R^s_{-s_4,-s_1}[\hat a_1,\hat a_2,\hat a_3,\hat a_4,\hat a] \hat p_4^{s_4}\hat p_1^{s_1}|a_i;a \rangle
\end{align}
By commuting with a B-move $\Omega_{12}^{a,\pm}$ we can easily derive simple operators $O^{s_1,s_3,\pm}_{1,3}$ and $O^{s_2,s_4,\pm}_{2,4}$ which join the corresponding punctures passing above or below the graph. 

With sufficient patience, one can verify the expected Weyl commutation relations between operators with the same set of $s_i$. 
A less trivial, but ultimately straightforward, task is to verify that these operators are compatible with the fusion and braiding transformations on
$\Gamma_0$. In other words, the know fusion integral kernel $F_{a,a'}[a_i]$ which maps s-channel and t-channel bases of conformal blocks into each other as in  
figure \ref{fig:4ptF} intertwines the corresponding expressions for the $O^{s_i,s_j}_{ij}$ operators acting on the two bases of conformal blocks. 
Conversely, such intertwining formulae can be interpreted as difference equations satisfied by the fusion and braiding kernels.
It would be interesting to pursue this line of inquiry further, possibly for other current algebras. 

We can do a few useful checks of the correctness of the expression for the $R^s_{s_2,s_3}$ coefficient. 
For example, if we set $a_2$ to the value for the identity, and $s_2=1$, 
$a = a_1 \pm \i b/2$, we can recover the appropriate triangle $T_{\pm,s_3}(a_1,a_3,a_4)$.
Also, we can compute the rectangle coefficient $R$ in an alternative way: fuse 
the two parallel horizontal rung and then compute the resulting triangles:
\begin{align}
R^s_{s_2,s_3}&[a_1,a_2,a_3,a_4,a]= I_s(a) T_{s_2,-s}[a_2,a+\i s b/2,a_1]T_{-s,s_3}[a+\i s b/2,a_3,a_4]
\end{align}

We can consider a double-rectangle, with $(2,1)$ rungs between the second and third legs of $\Gamma_0$
and compute it by first reducing the top rectangle and then the bottom, or viceversa. Concretely, 
that gives the identity 
\begin{align}
\sum_{s|s'-s = \pm1} &R^s_{s_2,s_3}[a_1+ \i s_1 b/2,a_2,a_3,a_4+ i s_4 b/2,a]R^{s'-s}_{s_4,s_1}[a_3,a_4,a_1,a_2,a+ \i s b/2] = \cr =
\sum_{s|s'-s = \pm1}  &R^{s'-s}_{s_2,s_3}[a_1,a_2,a_3,a_4,a+ \i s b/2]R^s_{s_4,s_1}[a_3+ \i s_3 b/2,a_4,a_1,a_2+\i s_2 b/2,a]
\end{align}
where $s'$ can be $2$,$0$,or $-2$.
This is useful in checking that $O^{s_2,s_3}_{2,3}$ and $O^{s_4,s_1}_{4,1}$ commute. 

We can also use the rectangle coefficient specialized to degenerate values of $a$ to fuse parallel
rungs in some generic graph. We will do so in a later section. Here we report the relevant coefficients:
if we specialize $a$ to a $(k+1,1)$ degenerate momentum, $a_1 = a_2 + \i (s_2 + s'_2) b/2$, 
$a_4 = a_3 + \i (s_3 + s'_3) b/2$, we get the coefficients
\begin{align}
C^{k;s}_{s_2,s_3;s'_2,s'_3}[a_2,a_3] &= s s_2 s_3 \frac{\alpha_2^{1-s} q^{s_2(k+2)/2+s'_2/2}-\alpha_2^{s-1} q^{-s_2(k+2)/2-s'_2/2}}{q^{k+1}-q^{-k-1}} \cdot \cr &\cdot\frac{\alpha_3^{1-s} q^{s_3(k+2)/2+s'_3/2}-\alpha_3^{s-1} q^{-s_3(k+2)/2-s'_3/2}}{q - q^{-1}}
\end{align}

\subsection{One-punctured torus}

\begin{figure}
\center \includegraphics[height=4.5cm]{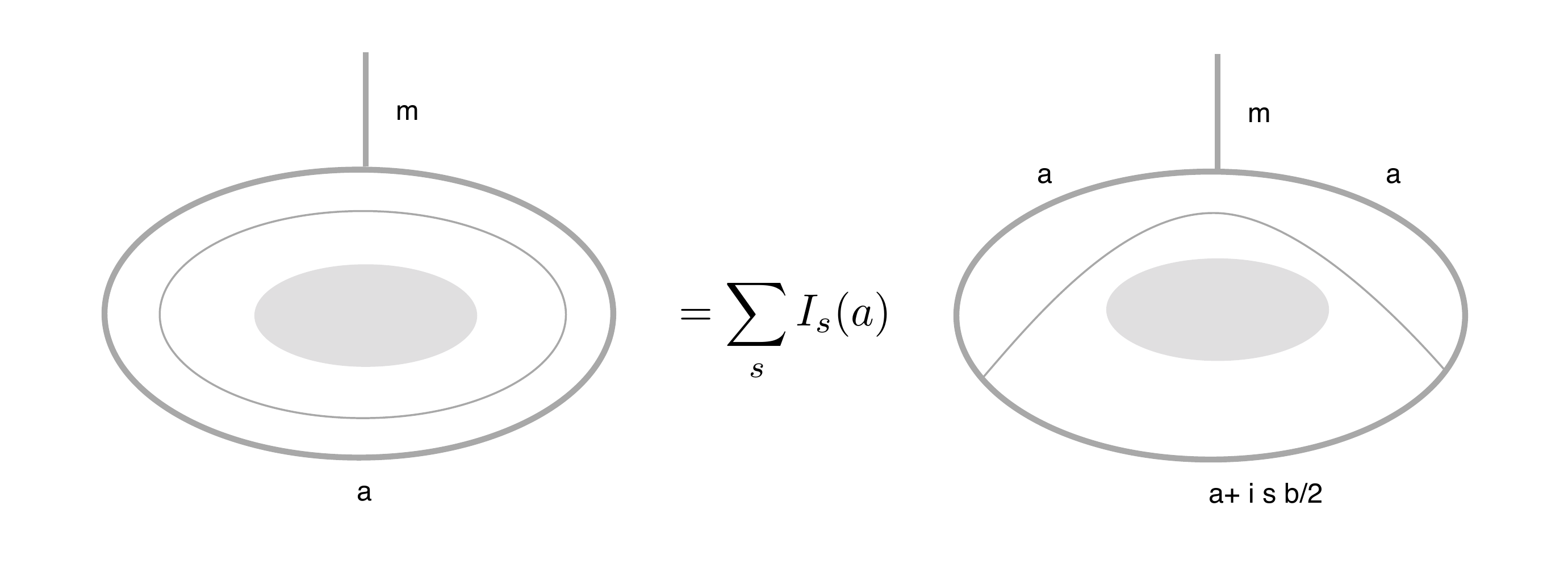}
\caption{\label{fig:bcycle} A closed Verlinde line operator wrapping a B-cycle inside the solid torus geometry associated to the one-punctured torus conformal block. The grey ellipse in the picture is a reminder that 
the loop cannot be shrunk inside the three-dimensional geometry. We depict the first step of the calculation, where an identity line is inserted between the B-cycle loop and the conformal block, and then subject to an A-move}
\end{figure}

The next classical example of closed Verlinde line operator wraps a B-cycle of the one-punctured torus. We will denote as $a$ the Liouville momentum running in the loop 
of the conformal block, and $m$ the Liouville momentum of the puncture ($\mu$ when exponentiated). We depict the cycle, and the first step of the calculation, in figure \ref{fig:bcycle}.

Inserting an identity line, and applying an A-move we arrive to a diagram we have already computed, with a degenerate line between two legs of a trinion. The final result is a linear combinations of conformal blocks with 
parameter $a + i s b/2$. We can write the B-cycle operator as 
\begin{equation}
\CL_{1,0} = \sum_s \i s p^s \frac{\mu \hat \alpha^{-2s} q^{-\frac{1}{2}} + \mu^{-1} \hat \alpha^{2s} q^{\frac{1}{2}} }{\hat \alpha^2 - \hat \alpha^{-2}} =  \sum_s \i s \frac{\mu \hat \alpha^{-2s} q^{\frac{1}{2}} + \mu^{-1} \hat \alpha^{2s} q^{-\frac{1}{2}} }{\hat \alpha^2 - \hat \alpha^{-2}}p^s 
\end{equation}
If we take the puncture to be the identity, $m = \i b/2 + \i b^{-1}/2$, the operator reduces to the expected $p + p^{-1}$.

We can easily compute variants of the B-cycle operator which wind $k$ times around the $a$ edge as they go around the torus, by conjugating the B-cycle operator 
with a $T^k$ operation on the torus, i.e. the scalar factor $e^{2 \pi \i k \Delta_a}$. Including an extra power of $q$ to untwist the ribbon of the degenerate line we get :
\begin{equation}
\CL_{1,k} = (-q^{1/2})^{-k} \sum_s \i s \hat \alpha^{2 s k} p^s \frac{\mu \hat \alpha^{-2s} q^{-\frac{1}{2}} + \mu^{-1} \hat \alpha^{2s} q^{\frac{1}{2}} }{\hat \alpha^2 - \hat \alpha^{-2}} 
\end{equation}

The A- and B- cycle closed Verlinde line operators obey nice algebraic relations which follow from skein relations. 
For example, 
\begin{equation}
\CL_{1,0} \CL_{0,1} = - q^{-1/2} \CL_{1,1}- q^{1/2} \CL_{1,-1}  \end{equation} 

There are two natural open line operators we will consider on the one-punctured torus. An ``A-cycle'' open operator looks like the loop in Figure \ref{fig:loopy} attached to the external leg of the one-punctured torus. 
\begin{equation}
{\cal O}_{0,1}^{s} = q  p_\mu^{2s} \frac{(\alpha^{1-s} \hat \mu q^{s/2} +\alpha^{-1+s} \hat \mu^{-1} q^{-s/2})(\alpha^{1-s} \hat \mu^{-1} q^{-s/2} +\alpha^{-1+s} \hat \mu q^{s/2}   )}{q - q^{-1}} 
\end{equation}
The formula is particularly simple for $s = 1$, where it reduces to 
\begin{equation}
-q  p_\mu^2 \frac{(\hat \mu q^{1/2} +\hat \mu^{-1} q^{-1/2})^2}{q - q^{-1}}
\end{equation}

\begin{figure}
\center \includegraphics[height=4.5cm]{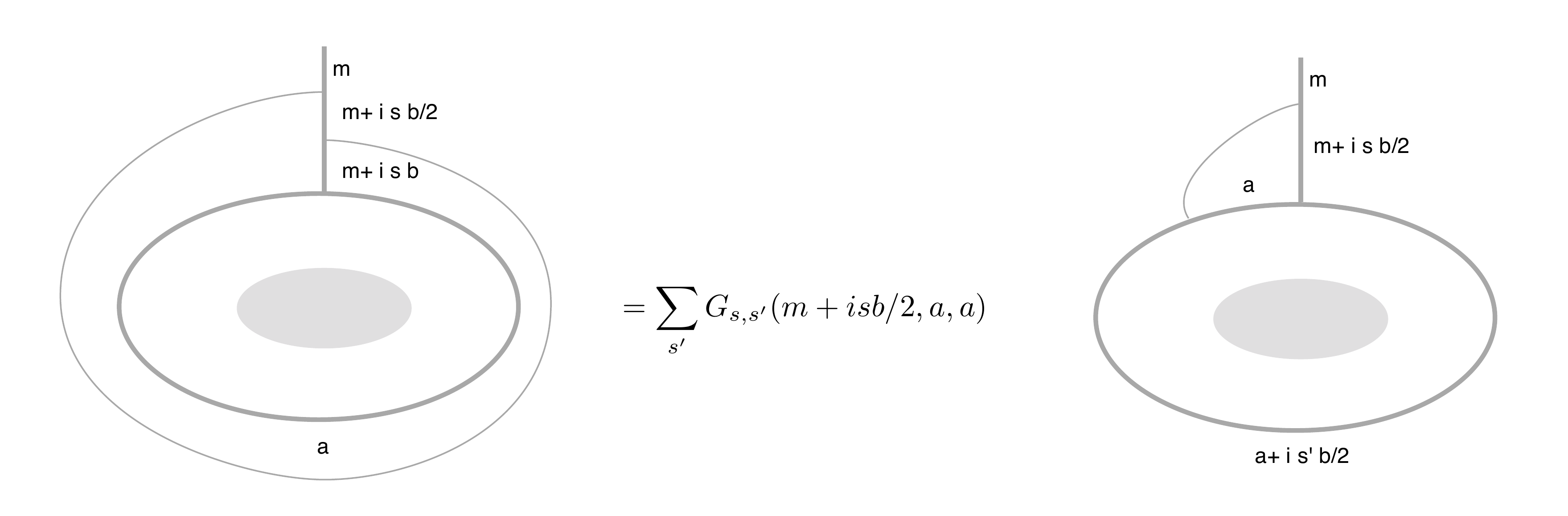}
\caption{\label{fig:openB} The computation of a ``B-cycle'' open Verlinde line defect for the one-punctured torus}
\end{figure}

A ``B-cycle'' open operator involves the calculation in figure \ref{fig:openB}

\section{Applications} \label{sec:app}
\subsection{Open Verlinde line operators and quantum Teichm\"uller theory}\label{sec:teich}
The algebraic relations satisfied by open and closed Verlinde line operators, such as the skein relations,
can be thought of as a quantum deformation of the algebraic relations satisfied by 
certain functions on the moduli space of (twisted) $\mathrm{SL}(2,C)$ flat connections over the Riemann surface:
traces of holonomies ${\mathrm Tr} \, \mathrm{Hol}_\ell$ and 
inner products between monodromy eigensections $(\psi^{s_j}_j, \mathrm{Hol}_{\ell_{ij}}\psi^{s_i}_i)$
at punctures. 

We can make the correspondence very concrete by looking at the semi-classical limit of conformal blocks with extra degenerate insertions. 
In the semiclassical limit $b \to 0$, keeping finite the classical Liouville momenta $A_i = a_i b$, the BPZ equations satisfied by the ``light'' degenerate fields $(2,1)$ in a conformal block reduce to a standard differential equation of the oper form 
\begin{equation}
\left[ \partial_z^2 + t(z,A_i) \right] \psi(z) =0
\end{equation} 
where $\psi(z)$ behaves as $dz^{1/2}$ under coordinate transformations, and $t(z,A_i)$ is a classical stress tensor which is such that the solutions have monodromies
$-e^{2\pi A_i}$ around the $i$-th leg of the conformal block. In particular, the solutions behave as $z^{1/2 \pm \i A}$ around a puncture of classical Liouville momentum $A$. 
Notice that the peculiar nature of $\psi(z)$, which is a section of the $K^{-1/2}$ bundle on the Riemann surface, 
means that some care is needed in defining its monodromies, i.e. one has to specify how the local coordinate system is transported. 
 
In general, that is why refer to twisted $\mathrm{SL}(2,C)$ flat connections.  
This means that one should think of parallel transport along ribbons. A twist of the ribbon can be undone at the price of a factor of $-1$.  This is the semiclassical limit of the usual $-q^{3/2}$.This is why, say, a closed contractible loop with a flat ribbon gives $-2$ rather than $2$.

Thus the semiclassical limit of a conformal block with an extra $(2,1)$ degenerate puncture at $z$ gives a flat section $\psi(z)$ of the oper. 
From the explicit form of the $B^\pm_s(a)$ matrix, we see that 
if the degenerate puncture is inserted on a specific edge $i$ and shifts the Liouville momentum by $\i b s/2$, the corresponding $\psi_{i,s}$ is an eigensection of the monodromy around that edge, with eigenvalues $- \alpha^{- 2 s}$. 

The semiclassical limit of the matrix elements which relate different bases of conformal blocks with the extra degenerate puncture placed on different edges give the matrix elements which relate the corresponding flat eigensections of the oper. Thus, for example, the holonomy of the oper, in a specific basis of eigensections, can be computed from the semiclassical limit of the sequence of $A$ and $B$ moves used to transport the degenerate puncture around the conformal block. 
Taking some care with following the semiclassical meaning of all intermediate steps in the definitions, it is easy to see that the 
 semiclassical limit of a closed Verlinde line operator is the trace of the holonomy of the oper along the corresponding path on the Riemann surface, while the semiclassical limit of an open line operator between two punctures is the Wronskian of two monodromy eigenvectors at the punctures, transported to a common point along the open path. More generally, the semiclassical limit of a general $|\Gamma\rangle$ evaluates come complicated network of Wilson lines for the oper, possibly ending on monodromy eigensections. 
  
The map from the space of opers in fixed complex structure to the space of twisted $\mathrm{SL}(2,C)$ flat connections over the Riemann surface
it not one-to-one: opers give a middle-dimensional, complex Lagrangian submanifold which is encoded in the semi-classical limit of conformal blocks: 
lhe conformal blocks behave as 
\begin{equation}
|a\rangle \sim e^{\frac{1}{b^2} {\cal W}[A]}
\end{equation}
and thus the quantum shift operator $\hat p^2$ becomes the complexified twist coordinates $p^2 = \exp \partial_A {\cal W}[A]$, which combine with the holonomies $\alpha = e^{ \pi A}$ to give complexified Fenchel-Nielsen coordinates. Indeed, in the semiclassical limit, the shift operators behave as $\hat p^2 = \exp b \partial_a \equiv \exp n b^2 \partial_A $.
The FN coordinate system is associated to a pair of pants decomposition of the Riemann surface, and the twist coordinate for a leg is 
defined by comparing two canonical normalizations for the monodromy eigensections for a leg of the pair of pants decomposition. 
The two canonical normalizations are associated to either of the trinions 
joined by the leg, and use the data of the monodromy eigensections at the other legs of each trinion. 
This matches well with the semiclassical limit of the degenerate field holonomy: the specific form of the $A$ move coefficients 
fix the relative normalization of all the eigensections for a trinion, and the comparison between nearby trinions involve the transport of the degenerate field 
along the leg, which shifts the Liouville momentum by $\pm b/2$ and rescales the flat sections by $p^{\pm 1}$.  

At least locally, if we allow the complex structure parameters to vary as well, we can parameterize the full space of flat connections.
Furthermore, before we act with the Verlinde line operators on the conformal blocks, 
they are given as a function of the A-cycle holonomies $A_i$ of the oper and their conjugate momenta $\log p_i$. 
The sequence of elementary moves we use to compute them provides a quantization of rules which work for any twisted $\mathrm{SL}(2,C)$ flat connection, 
not just for an oper. 

Thus we claim that the semiclassical limits of the operators $O_\ell$ and $O_{\ell_{12}}$ evaluate respectively the trace of the holonomy along a curve $\ell$ or the 
antisymmetric inner product of two monodromy eigensections compared along a path $\ell_{12}$ for a twisted flat connection labelled by 
the FN coordinates $A_i$ and $\log p_i$. When acting on conformal blocks, that specializes to the holonomy and Wronskians for the corresponding oper. 

The inner products $(\psi^{s_j}_j, \mathrm{Hol}_{\ell_{ij}}\psi^{s_i}_i)$ of two monodromy eigensections at punctures, 
specialized to the $\ell_{ij}$ being edges of an ideal triangulation and to a fixed choice of $s_i$,  are the building blocks for 
Fock/shear coordinates on Teichm\"uller space. More precisely, Fock coordinates are cross-ratios which do not depend on the choice of normalization of the eigensections. 
Thus we expect the open Verlinde line operators to be useful in understanding the relation between the quantization of 
Teichm\"uller space in FN coordinates and in Fock coordinates. 

Conventionally, the standard presentation of the quantization of Teichm\"uller space starts with Fock coordinates and uses them to 
reconstruct systematically the quantization of the traces of holonomies \cite{2005math}. Conformal blocks and quantum Teichm\"uller theory 
are related to each other through a Riemann-Hilbert problem \cite{Teschner:2003at,Vartanov:2013ima}. Our alternative approach based on open Verlinde line operators proceeds 
in the opposite direction, from conformal blocks and FN coordinates to quantum Teichm\"uller theory. 

It is easy to match our proposal for the quantum Wronskians and their cross-ratios with the explicit expressions 
computed for the one-punctured torus in \cite{Dimofte:2011jd}. It should be possible to also compare our results with the general integral kernels
constructed in \cite{Dimofte:2013lba}, which intertwine FN and Fock coordinates for the same Riemann surface. 

\subsection{Relation to $q$-deformed traffic rules and refined framed BPS degeneracies}
It is natural to conjecture, based on general supersymmetric gauge theory considerations, that the 
relation between Verlinde line operators and quantum Fock coordinates encodes the refinement of 
framed BPS degeneracies for theories in the class $A_1$, as defined and studied in \cite{Gaiotto:2010be}.

The framed BPS degeneracies should be encoded in the expansion of general Verlinde line operators
in terms of the collection of open Verlinde line operators 
\begin{equation}
\hat O_{ij} \equiv \hat O^{s_i,s_j}_{\ell_{ij}}
\end{equation}
associated to a choice of ideal triangulation with edges $\ell_{ij}$ and a fixed choice of shifts $s_i$, which picks a 
specific quantum eigensection. These are the quantum version of the objects denoted as ``$(s_i,s_j)$'' in that reference, 
where ``$s_i$'' denoted a choice of eigensection at each vertex.

In the cluster algebra language, the Fock coordinates/cross-ratios used in \cite{Gaiotto:2010be} are X-type variables, 
while the $O_{ij}$ operators are A-type coordinates. It is useful to cast our results in terms of the $O_{ij}$ rather than their cross-ratios. 
In particular, rather than expressing some operator $\hat O_\ell$ as a Laurent polynomial in the cross-ratios, we will 
write something like
\begin{equation}
\hat O_\ell \prod_{(i,j) | \ell_{ij} \cap \ell \neq 0} \hat O_{ij}^{\#(\ell_{ij} \cap \ell)} = \sum_{n_{ij}} c_{n_{ij}}[q] \prod_{i,j} \left( \hat O_{ij}\right)^{n_{ij}}
\end{equation}
Here $\ell$ can be a closed path, or any open path with $s_i$ labels at the ends. The $n_{ij}$ are non-negative powers, such that
$\sum_j n_{ij}$ equals the number of times operators of the type $\hat O_{i,\cdot}$ appear on the left hand side. The $c_{n_{ij}}[q]$
are Laurent polynomials in $q$. 

To get such an expression, we multiply $\hat O$ by an appropriate power of the operators associated to each of the edges $\ell_{ij}$ crossed by $\ell$.
That will allow us to use the skein relations to break $\ell$ into pieces, until each term of the sum only involves only a collection of paths 
associated to the edges of the triangulation. This can be done recursively, eliminating the intersections one-by-one
and breaking $\ell$ into smaller pieces.

The semi-classical limit of the skein operation applied to a product $\hat O_\ell \hat O_{ij}$ is very simple: it corresponds to inserting in the 
holonomy along $\ell$ a complete basis 
\begin{equation}
1 = \frac{(\cdot, \psi_i) (\psi_j,\cdot) - (\cdot,\psi_j) (\psi_i,\cdot)}{(\psi_i,\psi_j)}
\end{equation}
which is the basic step of the traffic rules discussed in \cite{Gaiotto:2010be}.

\subsection{Quantum groups}

There is a deep relation between conformal blocks and quantum groups. The most striking example is the fact that the fusion kernel for 
Virasoro conformal blocks coincides with the fusion kernel for the modular double quantum enveloping algebra 
${\cal U}_q(sl(2,R)) \times {\cal U}_{\tilde q}(sl(2,R))$, with $q = e^{i \pi b^2}$ and $\tilde q = e^{i \pi b^{-2}}$.
This relation and others involving Clebsh-Gordan coefficients is discussed in detail in \cite{2013arX}, which establishes a dictionary between the 
quantum groups and certain structures in quantum Teichm\"uller theory. 

In this section we would like to complement that analysis by attempting to reconstruct directly an action of the modular double quantum enveloping algebra
onto a certain space of conformal blocks. The quantum group generators in ${\cal U}_q(sl(2,R))$ will take the form of certain open Verlinde line operators 
labelled by the $(3,1)$ degenerate momentum, and the dual generators in ${\cal U}_q(sl(2,R))$ will take the form of certain open Verlinde line operators 
labelled by the $(1,3)$ degenerate momentum.

Our construction will be simply a re-formulation of a beautiful construction given in \cite{Witten:1989rw} in the context of Chern-Simons theory. 
In that reference, quantum group generators are built as operators which add rungs to webs of Wilson lines in Chern-Simons theory. 
The 3d TFT combinatorics of the construction is exactly the same as for open Verlinde line operators acting on conformal blocks,
and even the fusion and braiding coefficients will be closely related, as the ones for conformal blocks, at least 
if all momenta are specialized to degenerate values, are closely related to those for Wilson loops in Chern-Simons theory.

A crucial step in the construction in \cite{Witten:1989rw} actually works somewhat more naturally for conformal blocks: 
in order to recover the quantum group generators, the spin label of a certain reference line has to be analytically continued to infinity.
For conformal blocks, no analytic continuation is needed, as the momenta lie naturally in the continuum range. Furthermore, the limit 
as the momentum of an external leg is sent to infinity can be done in such a way that the standard space of conformal blocks is mapped to 
another well-defined space of conformal blocks: conformal blocks with an irregular singularity \cite{Gaiotto:2012sf}. 

More precisely, if we start from a regular conformal block, pick two external legs with momenta $a_2$ and $a_0-a_2$ 
and send $a_2$ to infinity while scaling the distance $z$ between the 
punctures to zero while keeping $\Lambda = a_2 z$ fixed, we obtain a conformal block with an irregular singularity of rank 1. 
The parameter $\Lambda$ behaves as a generalized complex structure parameter, while $a_0$ behaves much as a Liouville momentum. 

The BPZ equation has irregular singularities, and Stokes phenomena. One can still discuss the transport of degenerate fields 
around the Riemann surface, and define canonical bases of solutions of the BPZ equations with prescribed asymptotic behaviour at the irregular singularity,
which behaves in a similar fashion as the bases of monodromy eigenstates around regular punctures. 
Thus there are well-defined generalizations to the irregular case of standard conformal block notions, fusion and braiding operations, etc. 
Both closed and open Verlinde line operators should remain meaningful in the limit. Even operators which end on a leg whose momentum 
is sent to infinity, such as $O_k^{s_1,s_2}$, will have a meaningful limit. 

We leave the development of the full 3d TFT-like machinery for irregular conformal blocks to a separate publication. 
Here we will just focus on the algebraic aspects of the $a_2 \to \infty$ limit which are useful to make contact with the theory of quantum groups. 

We can look more closely at the algebra generated by open line defects stretched between two fixed edges. 
We can consider defects $O_k^{s_1,s_2}$ in the $(k+1,1)$ representation, which shift the Liouville momenta of the edges by $\i s_1 b/2$ and $\i s_2 b/2$.
These operators form a closed algebra under multiplication. Indeed, we can fuse the line defects with an A-move, and then eliminate the triangles. 

We already computed the structure constants involving $O_1$ and any $O_k$:
\begin{align}
\hat O^{s_1,s_2}_1 \hat O^{s'_1,s'_2}_k &= \sum_{s=\pm1} C^{k;s}_{-s_1,-s_2;-s'_1,-s'_2}[\hat a_1,\hat a_2] \hat O^{s_1 + s'_1,s_2+s'_2}_{k+s} \cr
\hat O^{s'_1,s'_2}_k \hat O^{s_1,s_2}_1  &= \sum_{s=\pm1} \hat O^{s_1 + s'_1,s_2+s'_2}_{k+s} C^{k;s}_{s_2,s_1;s'_2,s'_1}[\hat a_2,\hat a_1] 
\end{align}
In the above sums we should set to zero terms where $s_1 + s'_1$ or $s_2+s'_2$ lie outside the 
allowed range of shifts for the $\hat O_{k-1}$ operator.
Notice the relative order of coefficients and operators, which matter because $\hat \alpha_1 O_t^{s_1,s_2} = q^{s_1/2} O_t^{s_1,s_2}\hat \alpha_1$, etc. 

Remember the explicit expression 
\begin{align}
C^{k;s}_{s_2,s_3;s'_2,s'_3}[a_2,a_3] &= s s_2 s_3 \frac{\alpha_2^{1-s} q^{s_2(k+2)/2+s'_2/2}-\alpha_2^{s-1} q^{-s_2(k+2)/2-s'_2/2}}{q^{k+1}-q^{-k-1}}\cdot \cr &\cdot\frac{\alpha_3^{1-s} q^{s_3(k+2)/2+s'_3/2}-\alpha_3^{s-1} q^{-s_3(k+2)/2-s'_3/2}}{q - q^{-1}}
\end{align}
The two expressions match for $k=1$: 
\begin{align}
O_1^{s_1,s_2} O_1^{s_1',s_2'} = &(q + q^{-1})^{\delta_{s_1,s_1'}+ \delta_{s_2,s_2'}-1} O_2^{s_1 + s_1',s_2 + s_2'} +\cr -& \frac{\delta_{s_1+s_1',0}\delta_{s_2+s_2',0}}{q + q^{-1}}\frac{\hat \alpha_1^{2s_1} q - \hat \alpha_1^{-2 s_1} q^{-1}}{q - q^{-1}}\frac{\hat \alpha_2^{2s_2} q - \hat \alpha_2^{-2 s_2} q^{-1}}{q - q^{-1}}
\end{align}

In order to get a more standard algebraic structure, one could limit the $s_i$ shifts to be zero (and thus only use $(2k+1,1)$ defects). 
The problem is that the result is somewhat boring. The operators $O^{0,0}_{2k}$ can be written recursively as degree $k$ polynomials in $O^{0,0}_2$
and the algebra reduces to the algebra of polynomials in one variable. 

The next simplest case is to focus on operators with $s_1=0$. They commute with $\hat \alpha_1$. 
We can write a neat set of relations concerning the subset $O^{0,s_2}_2$. 
We can define $J^+ = (q - q^{-1}) O^{0,2}_2$ and $J^- =  (q - q^{-1})  O^{0,-2}_2$. Then we define 
\begin{align}
J^0 & \equiv  (\hat \alpha_2^2 - \hat \alpha_2^{-2})^{-1} (J^+ J^- - J^- J^+) = \cr &=(\alpha_1^2 + \alpha_1^{-2}) \frac{q - q^{-1}}{q + q^{-1}} O^{0,0}_2 
+\frac{(\alpha_1^2 q - \alpha_1^{-2} q^{-1})(\alpha_1^2 q^{-1} - \alpha_1^{-2} q) }{q^2 - q^{-2}}(\hat \alpha_2^2 + \hat \alpha_2^{-2})
\end{align}

We get interesting algebraic relations
\begin{align}
J^0 J^- - J^- J^0 &= (q + q^{-1}) (\hat \alpha_2^2 q- \hat \alpha_2^{-2} q^{-1}) J^- \cr
J^0 J^+ - J^+ J^0 &= (q + q^{-1}) (\hat \alpha_2^2 q^{-1}- \hat \alpha_2^{-2} q) J^+ \cr
\end{align}
We should also record the relation 
\begin{align}
 &(\hat \alpha_2^2 q- \hat \alpha_2^{-2} q^{-1}) O^{0,2}_2 O^{0,-2}_2 + (\hat \alpha_2^2 q^{-1}- \hat \alpha_2^{-2} q)O^{0,-2}_2 O^{0,2}_2 - \frac{ (\hat \alpha_2^2 - \hat \alpha_2^{-2})}{q + q^{-1}} O^{0,0}_2 O^{0,0}_2= \cr
 &=- (\hat \alpha_2^2 q- \hat \alpha_2^{-2} q^{-1})  (\hat \alpha_2^2 q^{-1}- \hat \alpha_2^{-2} q)  (\hat \alpha_2^2 - \hat \alpha_2^{-2}) \frac{( \alpha_1^2 q- \alpha_1^{-2} q^{-1})  (\alpha_1^2 q^{-1}- \alpha_1^{-2} q) }{(q + q^{-1})(q - q^{-1})^4}
\end{align}

We are ready to make contact with \cite{Witten:1989rw}: in order to simplify the algebra, we send $a_2$ to infinity. 

Our explicit examples and consistency of the composition of operators suggest setting 
$O_k = \hat \alpha_2^k o_k$ and keeping the $o_k$ finite in the limit. We thus define $J^a = \hat \alpha_2^2 t^a$ and
the relations 
\begin{align}
t^0 & \equiv q^{-2} t^+ t^- -  q^2 t^- t^+ = (\alpha_1^2 + \alpha_1^{-2}) \frac{q - q^{-1}}{q + q^{-1}} o^{0,0}_2 + \cr 
&+\frac{(\alpha_1^2 q - \alpha_1^{-2} q^{-1})(\alpha_1^2 q^{-1} - \alpha_1^{-2} q) }{q^2 - q^{-2}}
\end{align}
and
\begin{align}
q^{-1} t^0 t^- - q t^- t^0 &= (q + q^{-1}) t^- \cr
q t^0 t^+ - q^{-1} t^+ t^0 &= (q + q^{-1})t^+ \cr
\end{align}
and the Casimir-like expression
\begin{align}
 &q^{-1}o^{0,2}_2 o^{0,-2}_2 + q o^{0,-2}_2 o^{0,2}_2 - \frac{ 1}{q + q^{-1}} o^{0,0}_2 o^{0,0}_2= \cr
 &=- \frac{( \alpha_1^2 q- \alpha_1^{-2} q^{-1})  (\alpha_1^2 q^{-1}- \alpha_1^{-2} q) }{(q + q^{-1})(q - q^{-1})^4}
\end{align}
These are the quantum group relations found in \cite{Witten:1989rw}.

We can present some examples of quantum group representations which emerge from simple 
graphs $\Gamma_0$. This will also help in making contact with \cite{2013arX}.
First, we can take $\Gamma_0$ to be the trinion. In this special case we will have to also sent $\alpha_3$ to infinity in order 
to have a good irregular conformal block limit, keeping $\alpha_2 \alpha_3 = \alpha_0$ fixed. The result will be 
that our operators will act on irregular conformal blocks with one regular puncture and one irregular puncture. 

We have 
\begin{align}
O^{s_1,s_2}_1 =  -\i \frac{\hat \alpha_3 \hat \alpha_2^{-s_1} \hat \alpha_1^{-s_2} q^{\frac{s_1 s_2}{2}} + \hat \alpha_3^{-1} \hat \alpha_2^{s_1} \hat \alpha_1^{s_2} q^{\frac{-s_1 s_2}{2}} }{q - q^{-1}} \hat p_1^{s_1} \hat p_2^{s_2} 
\end{align}
We get 
\begin{align}
J^+ &= -\frac{(\hat \alpha_1 \hat \alpha_3^{-1} \hat \alpha_2^{-1}  q^{1/2} +\hat \alpha_1^{-1} \hat \alpha_3 \hat \alpha_2 q^{-1/2})(\hat \alpha_1 \hat \alpha_3^{-1}  \hat \alpha_2 q^{-1/2} +\hat \alpha_1^{-1} \hat \alpha_3 \hat \alpha_2^{-1}  q^{1/2})}{q - q^{-1}} p_2^{-2} \cr
J^- &= -\frac{(\hat \alpha_1 \hat \alpha_3 \hat \alpha_2 q^{1/2} +\hat \alpha_1^{-1} \hat \alpha_3^{-1} \hat \alpha_2^{-1} q^{-1/2})(\hat \alpha_1 \hat \alpha_3 \hat \alpha_2^{-1} q^{-1/2} +\hat \alpha_1^{-1} \hat \alpha_3^{-1} \hat \alpha_2 q^{1/2})}{q - q^{-1}} p_2^2 
\end{align}

Then we compute 
\begin{equation}
J^0 = -\frac{(\hat \alpha_1^2 + \hat \alpha_1^{-2})(\hat \alpha_3^2 + \hat \alpha_3^{-2})+(\hat \alpha_2^2 + \hat \alpha_2^{-2})(q + q^{-1})}{q - q^{-1}}
\end{equation}
which satisfies the desired commutation relations. 

Next, we take the limit $\alpha_2 \to \infty$ at fixed $\alpha_0 = \alpha_2 \alpha_3$:
\begin{align}
t^+ &= -\frac{ \alpha^2_1\hat \alpha_0^{-2}  + q^{-1} }{q - q^{-1}} p_0^{-2} \cr
t^- &= -\frac{q +\alpha_1^{-2} \hat\alpha_0^{-2}}{q - q^{-1}} p_0^2 \cr
t^0 &= -\frac{( \alpha_1^2 + \hat \alpha_1^{-2}) \hat\alpha_0^{-2}+q + q^{-1}}{q - q^{-1}}
\end{align}

We can make contact with a more familiar form of the quantum group algebra if we define 
\begin{align}
K &= \hat\alpha_0 \cr
E &= \hat\alpha_0 t^+ \cr
F &= \hat\alpha_0 t^- 
\end{align}

Indeed,
\begin{equation}
[E,F] = \hat\alpha_0^2 q^{-1} t^+ t^- - \hat\alpha_0^2 q t^- t^+  = -\frac{\hat\alpha_0^2 -\hat\alpha_0^{-2}}{q - q^{-1}}
\end{equation}
This is basically the same as the basic quantum group representation discussed in \cite{2013arX}.

If we repeat the exercise for more interesting graphs $\Gamma_0$, 
we will obtain several distinct simultaneous actions of quantum group generators on the space of conformal blocks with an irregular puncture and $n$ 
regular punctures. Each action will be labelled by one of the regular punctures, and a path from the irregular puncture to the regular punctures. 

Different sets of quantum group generators associated to paths which only intersect at the irregular puncture 
will have some twisted commutation relations governed by the $\alpha_2 \to \infty$ of the
braiding matrix for two $(3,1)$ insertions on an edge of momentum $a_2$.  

Different bases of irregular conformal blocks will give different representations of the quantum group generators, 
intertwined by the fusion and braiding kernels for conformal blocks. With some extra work, it should be possible to 
identify some of these representations with the twisted tensor product representations employed in \cite{2013arX},
and thus show the match between conformal block and quantum group fusion kernels and R-matrices.

It would also be interesting to explore the representations associated to punctures on a higher genus Riemann surface.

\section*{Acknowledgements}
DG thanks Joerg Teschner for discussions and comments on the draft. 
Much of this work was developed as a result of useful questions, discussions and suggestions by Jaume Gomis, Bruno Le Floch and Edward Witten. 
We also thank Alexander Braverman for rekindling DG interest in this subject. The research of DG was supported by the Perimeter Institute for Theoretical Physics. Research at Perimeter Institute is supported by the Government of Canada through Industry Canada and by the Province of Ontario through the Ministry of Economic Development and Innovation. 

\bibliographystyle{JHEP}

\bibliography{openloops}

\end{document}